\documentclass[12pt]{article}
\usepackage{amsmath,amssymb,amsfonts,amsthm,bbm}
\usepackage{epic,eepic,epsfig,longtable}
\usepackage{multirow,verbatim}
\usepackage{array}
\usepackage{graphicx}
\usepackage{floatrow}
\usepackage{subcaption}
\usepackage{paralist}
\usepackage{latexsym}
\usepackage{comment}
\usepackage{epsfig}
\usepackage{setspace}
\usepackage{enumitem}
\usepackage{CJK}
\usepackage{color}
\usepackage{booktabs}
\usepackage{rotating}
\usepackage[authoryear, round]{natbib}
\newcommand\numberthis{\addtocounter{equation}{1}\tag{\theequation}}

\makeatletter
\def\singlespace{\def\baselinestretch{1}\@normalsize}

\makeatletter
\def\singlespace{\def\baselinestretch{1}\@normalsize}

\renewcommand{\theequation}{\thesection.\arabic{equation}}
\numberwithin{equation}{section}

\renewcommand{\hat}{\widehat}

\renewcommand{\hat}{\widehat}

\newcommand{\bfm}[1]{\ensuremath{\mathbf{#1}}}

    \def\FF{\mathbb{F}}

\def\bx{\bfm x}

\newcommand{\bfsym}[1]{\ensuremath{\boldsymbol{#1}}}

\def\1{\bfsym{1}}	
\def\bbE{\mathbb{E}}

\DeclareMathOperator{\var}{var}

\def\var{\mbox{var}}

\def\today{\ifcase\month\or
  January\or February\or March\or April\or May\or June\or
  July\or August\or September\or October\or November\or December\fi
  \space\number\day, \number\year}

\newdimen\biblioindent    \biblioindent=30pt

 at 8truept

\newcommand{\beq}{\begin{equation}}
  \newcommand{\eeq}{\end{equation}}
\newcommand{\beqn}{\begin{eqnarray}}
  \newcommand{\eeqn}{\end{eqnarray}}
\newcommand{\beqnn}{\begin{eqnarray*}}
  \newcommand{\eeqnn}{\end{eqnarray*}}

\allowdisplaybreaks
\def\thesection{\arabic{section}}
\usepackage{verbatim}
\def\FF{\mathcal{F}}
\def\[{\left [}  \def\]{\right ]} \def\({\left (}  \def\){\right )}
 \def\endpf{$\blacksquare$}
\def\hat{\widehat}
\makeatletter
\def\underbar#1{\underline{\sbox\tw@{$#1$}\dp\tw@\z@\box\tw@}}
\makeatother

\newtheorem{assumption}{Assumption}
\newtheorem{theorem}{Theorem}
\newtheorem{lemma}{Lemma}

\theoremstyle{definition}
\newtheorem{definition}{Definition}

\newtheorem{remark}{Remark}

\title{Exponential GARCH-It\^o Volatility Models}
\author{Donggyu Kim   \\
College of Business, \\
Korea Advanced Institute of Science and Technology (KAIST)
}

\begin{document}
\maketitle

\begin{spacing}{1.45}

\begin{abstract}
This paper introduces a novel It\^o diffusion process to model high-frequency financial data, which can accommodate low-frequency volatility dynamics by embedding the discrete-time non-linear exponential GARCH structure with log-integrated volatility in a continuous instantaneous volatility process. 
The key feature of the proposed model is that, unlike existing GARCH-It\^o models, the instantaneous volatility process has a non-linear structure, which ensures that the log-integrated volatilities have the realized GARCH structure.
We call this the exponential realized GARCH-It\^o (ERGI) model. 
Given the auto-regressive structure of the log-integrated volatility, we propose a quasi-likelihood estimation procedure for parameter estimation and establish its asymptotic properties.
We conduct a simulation study to check the finite sample performance of the proposed model and an empirical study with 50 assets among the S\&P 500 compositions. 
The numerical studies show the advantages of the new proposed model. 

\end{abstract}



\section{Introduction} \label{SEC-1}

\begin{figure}[!ht]
  \centering
    \includegraphics[width=0.9\textwidth]{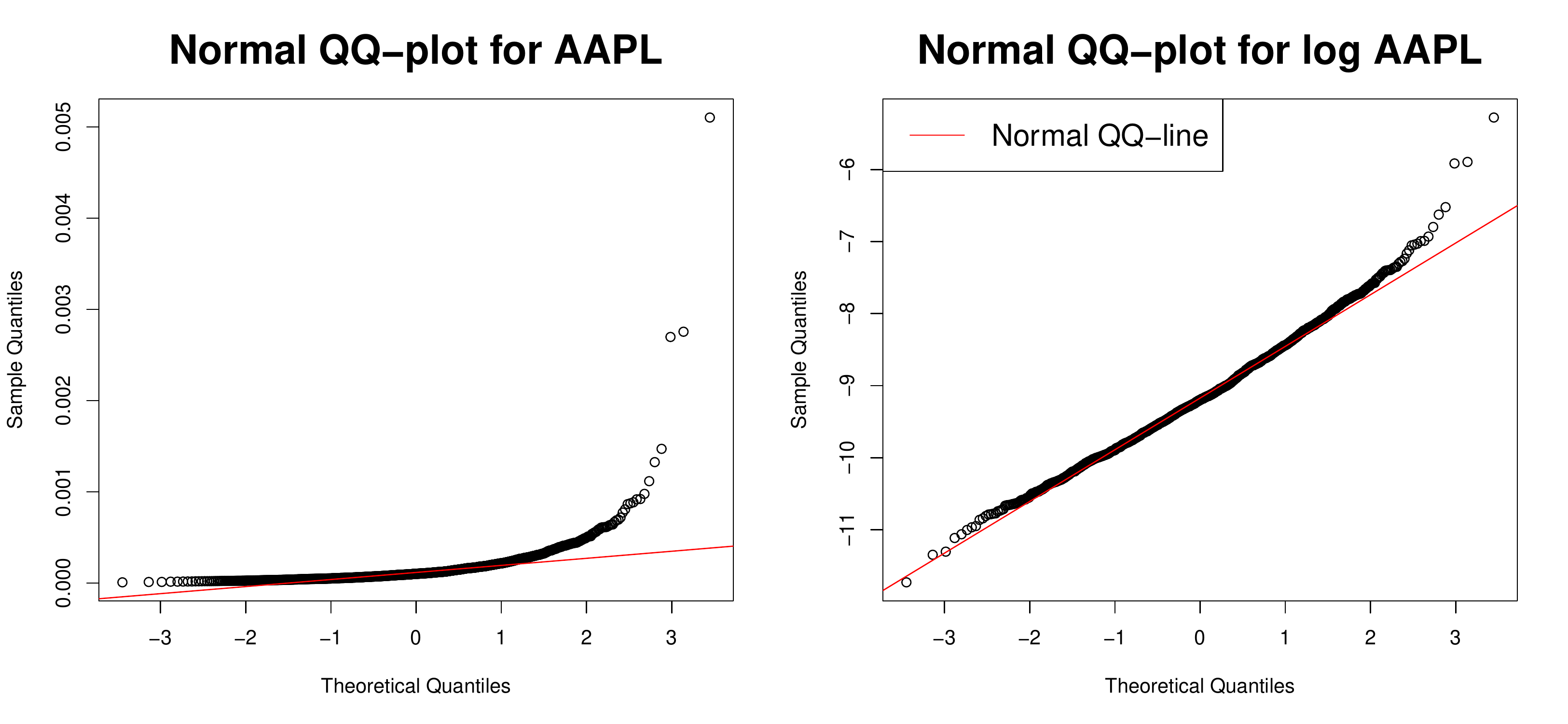}
     \caption{Normal QQ-plots of realized volatilities and log realized volatilities for AAPL. 
     The red real line denotes the best linear fit line, which illustrates perfect normal distribution.}
     \label{Figure-NormalQQ}
\end{figure}

 In finance practice, volatility plays a  pivotal role.
Low-frequency and high-frequency financial data are widely used to analyze  volatility dynamics.
 For example, generalized auto-regressive conditional heteroskedasticity (GARCH) models are introduced to catch the low-frequency volatility dynamics, such as volatility clustering, by employing the squared low-frequency log-return as the innovation \citep{bollerslev1986generalized, engle1982autoregressive}. 
However, when the volatility changes rapidly, it is often difficult to catch  the change  using only the low-frequency log-returns as the innovations \citep{andersen2003modeling}.
  On the other hand, high-frequency financial data are available to construct the so-called realized volatility for estimating daily integrated volatility.
Examples include  two-time scale realized volatility (TSRV) \citep{zhang2005tale}, multi-scale realized volatility (MSRV) \citep{zhang2006efficient}, kernel realized volatility (KRV) \citep{barndorff2008designing},
quasi-maximum likelihood estimator (QMLE) \citep{ait2010high, xiu2010quasi}, pre-averaging realized volatility (PRV) \citep{jacod2009microstructure}, and robust pre-averaging realized volatility \citep{fan2018robust, shin2021adaptive}.
These realized volatility estimators contain high-frequency information about the market volatility, and many studies show that incorporating high-frequency information   helps  account for  low-frequency market dynamics \citep{corsi2009simple, hansen2012realized, kim2016unified, shephard2010realising}. 
 Several conditional volatility models have been developed to combine high-frequency and low-frequency data and enhance volatility estimation and  predication by employing realized volatility as the volatility proxy.
 Examples include the realized volatility based modeling approaches 
\citep{andersen1997heterogeneous, andersen1997intra-day, andersen1998skeptics, andersen1998deutsche,  andersen2003modeling}, 
the heterogeneous auto-regressive (HAR) models \citep{corsi2009simple}, the realized GARCH models \citep{hansen2012realized},  the high-frequency based volatility (HEAVY) models \citep{shephard2010realising},  and the unified GARCH/SV-It\^o models \citep{kim2019factor, kim2016unified, song2021volatility}.
These models have been developed based on the linear auto-regressive structure of realized volatilities. 
However,  we often observe that  non-linear auto-regressive structures, such as exponential functions, better capture the volatility dynamics \citep{nelson1991conditional, hansen2016exponential, kawakatsu2006matrix}.
This may be because log-volatilities often have a stronger linear auto-regressive relationship.
In fact, when variables are close to normal distributions,  linear models work well. 
To check the normality of realized volatilities, we draw normal QQ-plots of realized volatilities and log realized volatilities for AAPL stock.
Figure \ref{Figure-NormalQQ} shows that the log-transformation makes the realized volatilities closer to a normal distribution. 
Most of assets have the similar phenomena. 
Thus, we can conjecture that log-volatilities better explain volatility dynamics. 
To harness this feature, \citet{hansen2016exponential} employed the exponential GARCH structure with the log-realized volatility as the innovation, and their empirical study indicates that the non-linear GARCH structure helps account for market dynamics. 
Although, as discussed above, empirical studies support that incorporating high-frequency data with a non-linear auto-regressive structure better captures the market dynamics, the mathematical gap between the empirical low-frequency discrete-time non-linear volatility models, such as the exponential realized GARCH, and high-frequency-based continuous-time diffusion process is not well-studied. 
In fact,  several studies have been conducted to fill the gap between the discrete-time volatility models and continuous-time diffusion process \citep{kim2019factor, kim2016unified, song2021volatility}.
However, these studies are based on the linear auto-regressive structure, and the extension from linear to non-linear structures is not straightforward.
This fact increases the demand for developing continuous-time diffusion process-based models that provide a rigorous mathematical 
formulation for the non-linear auto-regressive structure of realized volatilities.

In this paper, we develop a novel diffusion process to model high-frequency financial data, which can accommodate a non-linear GARCH structure of the realized volatilities. 
From the empirical study, we often observe that the log-realized volatility has a stronger auto-regressive structure than the original realized volatility. 
To reflect this, we employ the exponential GARCH structure as the non-linear function.
Specifically, the log-realized volatility follows the realized GARCH structure. 
To connect this low-frequency volatility structure with the continuous-time diffusion process, we develop a novel continuous instantaneous volatility process. 
Since the  volatility process has a non-linear structure, the linear structure of the unified GARCH-It\^o \citep{kim2016unified} is not applicable. 
Furthermore, usual log-diffusion processes for instantaneous volatility processes can not provide the solution.
To tackle this issue, we propose a novel instantaneous volatility process, based on the average integrated volatility process.
Then, the proposed instantaneous volatility process is continuous with respect to time, and its daily integrated volatility is decomposed into the exponential realized GARCH with the daily log-integrated volatility as the innovation and exponential martingale difference. 
We call it the exponential realized GARCH-It\^o (ERGI) model.  
Unlike the linear realized GARCH model, the log-realized volatility can have negative values.
Thus, we allow model parameters to be negative. 
To estimate the model parameter, we propose a quasi-maximum likelihood estimation procedure.
Specifically, we adopt the Gaussian quasi-likelihood function and use the realized volatility as the proxy of the conditional expected value. 
Furthermore, we establish its asymptotic properties. 
To illustrate the benefit of the proposed model, we apply the ERGI model to real trading high-frequency data and find that the exponential structure helps  account for the volatility dynamics.

The rest of paper is organized as follows.
In Section \ref{SEC-2}, we propose the ERGI model and investigate its properties.
In Section \ref{SEC-EST}, we suggest the quasi-maximum likelihood estimation procedure and study its asymptotic behaviors.
In Section \ref{SEC-simulation}, we conduct a simulation study to check the finite sample performance of the ERGI model. 
In Section \ref{SEC-empiric}, we apply the ERGI model to the top 50 high trading volume assets among the S\&P 500 compositions. 
Section \ref{SEC-conclude} contains the conclusions.
 All the technical proofs are collected in Section \ref{SEC-proof}.

\setcounter{equation}{0}

\section{Exponential realized GARCH-It\^o models} \label{SEC-2}

In this section,  we develop an exponential realized GARCH-It\^o (ERGI) model as follows.
\begin{definition}\label{Def-1}
We call the log-price $X_t$ to follow the  ERGI process if it satisfies
\begin{eqnarray*}
	&&dX_t= \mu_t dt + \sigma_t (\theta) dB_t + J_t d \Lambda_t , \cr
	&&\sigma_{t}^2 (\theta)   = \bar{\sigma}_t^2(\theta) \{ 1+(t-[t]) b_t (\theta) \}, 	\cr
	&&b_t (\theta) =  b_{[t]}(\theta) + (  t- [t])     ( \omega  + ( \gamma-1)   b_{[t]} (\theta)    )   +  \beta  \log \bar{\sigma}_t^2(\theta) \cr
    && \qquad  \qquad -(1 - t +[t]) \left( \beta  +  \beta^*  ( t-[t])  \right) \log \sigma_{[t]}^2  +   \nu  (1- t +[t])    Z _t ^2 ,   
\end{eqnarray*}
 where $\bar{\sigma}_t^2 (\theta) = (t-[t])^{-1}\int_{[t]}^t \sigma^2 _s (\theta) ds$,  
 $[t]$ denotes the integer part of $t$, except that $[t]=t-1$ when $t$ is an integer,    $Z_t  = \int_{[t]}^t   dW_s$,  and $\theta = (\omega,  \gamma,  \beta, \beta^*, \nu )$ is the model parameter. 
 For the jump part, $J_t$ is the jump size and $\Lambda_t$ is the Poisson process with the intensity $\lambda_t$. 
\end{definition}

We note that $\beta^*$ is an adjust term to handle some remaining terms. 
Specifically, we set $\beta^*  = \frac{1+\beta \varrho_2 }{ \varrho_2 - 2 \varrho_3}$, where $\varrho$'s are functions of $\beta$, which are defined in Theorem \ref{thm-integratedVol}.
The log-average integrated volatility, $\log \bar{\sigma}_t^2 (\theta)$, provides the innovation term, and  $Z_t$  is the random fluctuation.
Since the process $b_t(\theta)$ is continuous with respect to time $t$, the instantaneous volatility process $\sigma_t^2 (\theta)$ is also continuous.
At the integer time points, we have
 \begin{equation*}
 	b_n  (\theta) = \omega + \gamma b_{n-1} (\theta) +   \beta  \log  \int_{n-1} ^{n} \sigma_t^2 (\theta) dt .
 \end{equation*}
  That is, $b_n (\theta)$ can be explicitly expressed by the past log-integrated volatilities, and $b_t (\theta)$ has a form of the interpolation between these values.
 Thus, when considering $b_t(\theta)$, the ERGI model has a similar structure of realized GARCH-It\^o model \citep{song2021volatility} with the log-integrated volatility as the innovations. 
 However, unlike the realized GARCH-It\^o model, to obtain the non-linear exponential realized GARCH form, the instantaneous volatility process has a non-linear structure, such as $\bar{\sigma}_t^2(\theta) \{ 1+(t-[t]) b_t (\theta) \}$. 
The solution for this structure is 
 $$
 	\frac{1}{t-n+1}\int_{n-1}^t \sigma_s^2 (\theta) ds = \sigma_{n-1}^2(\theta) e^{\int_{n-1}^t b_s(\theta) ds} \text{ a.s.}
$$
Details can be found in Lemma \ref{exp-lemma}.
   Using the above solution, we can measure the integrated volatility $\int_{n-1}^n \sigma_t^2(\theta) dt$ with $\sigma_{n-1}^2(\theta)$ and $\int_{n-1}^n b_s(\theta) ds$, which has the realized GARCH form with the log-integrated volatilities.
Specific properties of the integrated volatility are shown in the following theorem.

 \begin{theorem} 	\label{thm-integratedVol}
Under the ERGI model, for $|\beta|<1$ and $ n \in \mathbb{N}$,  the integrated volatilities have the following properties.
	\begin{enumerate}
		\item[(a)] We have
			\begin{eqnarray*}
			&&\int_{n-1} ^{n}  \sigma_{t}^2(\theta)  dt  = \exp \left( h_n(\theta)   + D_n \right) \quad \text{a.s.}, 	\cr
	&&h_n(\theta)  = \omega^*  + \gamma h_{n-1}(\theta) + \beta^g \log \int_{n-2}^{n-1}\sigma_t^2(\theta)dt  ,
			\end{eqnarray*}
		where    
	\begin{eqnarray*}
	&&\omega^{*}   =    \{ (1-\gamma)  \varrho_2  + \varrho \}  \omega  + (1-\gamma) \nu (\varrho_2 - 2 \varrho_3)  , \quad  \beta^g = \varrho \beta, \quad   \varrho = \varrho_{1} +(\gamma-1 )  \varrho_{2},    \cr
	&&  \varrho_{1} = \beta ^{-1} ( e^{\beta} -1), \quad  \varrho_{2} = \beta ^{-2} ( e^{\beta } -1 - \beta), \quad  \varrho_{3} = \beta ^{-3} ( e^{\beta } -1 - \beta  -   \beta  ^2/2  ), 
\end{eqnarray*}
		and
			\begin{eqnarray*}
			&& D_n = 2 \nu   \int_{n-1}^n \{  ( n-t) \beta^{-1}  e^{\beta (n-t)}   - (e^{\beta (n-t)} -1   ) \beta^{-2}  \}Z_t dW_t 
			\end{eqnarray*}
		is a martingale difference.

			\item[(b)]We have
\begin{eqnarray}
&&\int_{n-1} ^{n}  \sigma_{t}^2(\theta)  dt = \exp \left( H_n(\theta)   \)  M_n , \cr
&&    \bbE \left[ \int_{n-1}^{n}  \sigma^2_t (\theta) dt \middle| \mathcal{F}_{n-1} \right]  =  \exp \left( H_n (\theta)   \right) \text{ a.s.},   \label{r3-Thm1}
\end{eqnarray}		
where 
\begin{eqnarray*}
&&H_n (\theta) = \omega^g + \gamma H_{n-1}(\theta) + \beta^g \log \ \int_{n-2}^{n-1}\sigma_t^2(\theta)dt  ,  	\cr
&&\omega^g = \omega^* +  (1-\gamma)    \log \bbE [ \exp (D_n) ],
\end{eqnarray*} 
 and
\begin{equation*}
	M_n = \exp ( D_n -  \log \bbE [ \exp (D_n) ])
\end{equation*}
is an exponential martingale difference. 
	\end{enumerate}
\end{theorem}

Theorem \ref{thm-integratedVol}(a) shows that the log-integrated volatility is decomposed into the realized GARCH with the log-integrated volatility innovations, $h_n(\theta)$, and the martingale difference $D_n$. 
Thus, the log-realized GARCH, $h_n(\theta)$, is the conditional expected value of the log-integrated volatility, but it is not the conditional expected value of the original integrated volatility. 
 In Theorem \ref{thm-integratedVol}(b), we show that the integrated volatility is decomposed into the exponential function of the realized GARCH with the log-integrated volatility innovations, $H_n(\theta)$, which has the additional interceptor term from the martingale difference term $D_n$, and the exponential martingale difference $M_n$.
 Since the expectation is a linear function, the additional  interceptor term does not appear in the linear realized GARCH.
 However, the ERGI is non-linear, and, thus,  we have the additional  interceptor term.
 The main purpose of this paper is to develop a model for analyzing the original integrated volatility, so we develop a statistical inference based on \eqref{r3-Thm1}.
Theorem  \ref{thm-integratedVol} indicates that the proposed model has a non-linear exponential GARCH structure.
From the empirical study, we find that this non-linear structure helps  explain the volatility dynamics as compared to the usual linear realized GARCH. 
Details can be found in  Section \ref{SEC-empiric}.

 \subsection{Relationship with the daily log-returns}

The traditional discrete GARCH models are models of close-to-close volatilities for log-returns.
In this section, we discuss the relationship between the proposed ERGI and close-to-close volatilities for log-returns. 

We first consider the continuous part.
That is, we assume that the log-price process does not have the jump part. 
Then, by It\^o's lemma and Theorem \ref{thm-integratedVol}(b), we have
\begin{eqnarray*}
\bbE  \[ \(X_n - X_{n-1} -\int_{n-1}^n \mu_t dt    \)^2 \middle| \FF_{n-1} \] &=& \bbE  \[  \( \int_{n-1}^n \sigma_t (\theta) dB_t\)^2 \middle | F_{n-1}\]   \cr	 
		&=&  \exp \left( H_n (\theta)   \right)     \text{ a.s.} 
\end{eqnarray*}
Thus, the exponential GARCH volatility, $\exp \left( H_n (\theta)   \right)$, is the conditional volatility of the daily log-return.  
Unfortunately, in practice, we do not have the observations during the close-to-open period. 
Thus, in order to investigate the close-to-close volatility, we need to impose a structure on the overnight period. 
For example, we can simply use squared close-to-open log-returns as as the proxy of the integrated volatility for the close-to-open period. 
Then, we can apply the proposed ERGI model to the realized volatility plus squared close-to-open log-return. 
 On the other hand, we can assume that the close-to-open volatility dynamics is same as the open-to-close period.
 Then, we only need to match the scale.
 To do this, we can calculate the averages for the open-to-close realized volatilities and the squared close-to-open log-returns, and we multiply the inverse of the proportion of the average of the realized volatility.  
 The above methods are practical solutions without theoretical justifications. 
 Thus, it is interesting and important to develop a diffusion process which can accommodate the close-to-close period.
 We leave this for future study.

To investigate the jump diffusion process. 
We assume that the jump sizes $J_t$'s are i.i.d. with mean $\mu_J$ and variance $\sigma_J^2$, and the intensity is constant over time; that is, $\lambda_t=\lambda$. 
Furthermore, we assume the continuous part and jump part are not correlated. 
Then, the conditional volatility of the daily log-return is
\begin{eqnarray*}
	 \bbE  \[ \(X_n - X_{n-1} -\int_{n-1}^n \mu_t dt    \)^2 \middle| \FF_{n-1} \]  =  \exp \left( H_n (\theta)   \right)    + \lambda \sigma_J^2 +  \lambda^2 \mu_J ^2 \text{ a.s.} 
\end{eqnarray*}
The squared log-return has the  exponential GARCH, $\exp \left( H_n (\theta)   \right)$, and additional expected jump variation.
The jump variation part depends on the assumption of the jump structure. 
Thus, it is interesting and important to investigate the jump variation dynamics and to model the jump part.
 We leave this for future study.

 \section{Estimation procedure} \label{SEC-EST}
  
 \subsection{A model setup }
  We assume that the log-prices follow the ERGI process defined in Definition \ref{Def-1}. 
  The intra-day log-prices for the $d$th day are observed at $t_{d,i}, i=1,\ldots, m_d$, where $d-1=t_{d,0} < t_{d,1} < \cdots < t_{d, m_d} = 1+ d-1$.  
  We denote $m$ as the average number of the high-frequency observations; that is, $m= \frac{1}{n} \sum_{d=1}^n m_d$.  
Unfortunately, true high-frequency observations,  $X_{t_{d,i}}$'s, are not observed due to market micro-structure noises. 
To accommodate the market micro-structure noises, we assume that the observed log-prices $Y_{t_{d,i}}$ has the following structure:
 \begin{equation*}
 	Y_{t_{d,i}}= X_{t_{d,i}} + \epsilon_{t_{d,i}}, \quad \text{for } d=1,\ldots, n, i=1,\ldots, m_d, 
 \end{equation*}
 where $X_t$ is the true log-price and $\epsilon_{t_{d,i}}$'s are micro-structure noises with mean zero.

Without the presence of price jumps,  several nonparametric realized volatility estimators have been constructed that take advantage of sub-sampling and local-averaging techniques to remove the effect of market micro-structure noises so that the integrated volatility can be estimated consistently and efficiently \citep{barndorff2008designing, fan2018robust, jacod2009microstructure, shin2021adaptive, zhang2006efficient, xiu2010quasi}. 
To identify the jump locations given noisy high-frequency data, \citet{fan2007multi} and \citet{zhang2016jump} proposed wavelet methods to detect jumps and applied the MSRV method to jump-adjusted data. 
They showed that the estimator of jump variation has the convergence rate of $m^{-1/4}$,  and  the estimator of integrated volatility achieves the optimal convergence rate of $m^{-1/4}$. 
On the other hand,  \citet{ait2016increased} proposed jump robust pre-averaging methods by employing a truncation method.  
They also demonstrated that the estimators of jump variation and integrated volatility achieve the optimal convergence rate of $m^{-1/4}$. 
In this paper, for the $i$th day, we let $RV_i$ to be the corresponding estimator of daily integrated volatility that is robust to micro-structure noises and price jumps.
In the numerical study, we employ the jump robust pre-averaging method.

 \subsection{GARCH parameters estimation}\label{Asymptotics}
 We first fix notations. 
 For a given vector $\bx = (x_i)_{i=1,\ldots,k}$, we define $\| \bx \|_{\max} = \max_i |x_i|$.
 Let $C$'s be generic constants whose values does not depend on $\theta^g, n$, and $m$ and may change from occurrence to occurrence.
 In this section, we develop an estimation procedure for  the true GARCH model parameters $\theta^g_0 = (\omega^g_0, \gamma_0, \alpha^g_0)$.

 Theorem \ref{thm-integratedVol} indicates that the integrated volatility is decomposed into the exponential GARCH term $\exp \left(H_n(\theta^g) \right)$ and the exponential martingale difference term $M_n$, which implies 
\begin{equation*}
\frac{\int_{n-1} ^{n}  \sigma_{t}^2(\theta_0^g )  dt-  \exp \left( H_n(\theta^g_0 )   \) }{   \exp \left( H_n(\theta^g_0)   \) } =  M_n -1 \text{ a.s.}
\end{equation*}
Since $M_n$ is the exponential martingale difference, $M_n-1$ is a martingale difference.
 This   inspires us to use integrated volatility as a proxy for exponential GARCH volatility, and  we define a quasi-likelihood function as follows:
\begin{equation*}
	\hat{L}_n(\theta^g) = - \frac{1}{n}\sum_{i=1}^n \left\{H_i(\theta^g) + \frac{ \int_{i-1}^i \sigma_t^2(\theta^g_0) dt}{\exp\left( H_i(\theta^g) \right)}\right\}.
\end{equation*} 
We can estimate the parameter $\theta^g$ by maximizing the above quasi-likelihood function.
   However, in practice, the integrated volatility is not observable, so we need to estimate it first.
We employ the jump robust realized volatility estimator  \citep{ait2016increased, fan2007multi, zhang2016jump}.
 Then, we estimate the log-conditional expectation of the integrated volatilities as follows:
 \begin{equation}\label{eq-Hhat}
 \hat{H}_i(\theta^g) = \omega^g + \gamma \hat{H}_{i-1}(\theta^g) + \beta^g \log RV_{i-1},
 \end{equation}
 where the initial value $\hat{H}_1(\theta^g)$ is set to be $\log RV_1$.
 The effect of the initial value is negligible with the rate of $n^{-1}$ (see Lemma 1 in \citet{kim2016unified}), so its choice does not have a significant effect on the parameter estimation.
 With the estimated conditional expected volatility function, we define the following quasi-likelihood function:
 \begin{equation*}
 \hat{L}_{n,m}(\theta^g) =- \frac{1}{n}\sum_{i=1}^n \left\{\hat{H}_i(\theta^g) + \frac{ RV_{i}   }{\exp\left( \hat{H}_i(\theta^g) \right)}\right\}.
	\end{equation*}
Then, we obtain the estimator for the GARCH parameters $\theta^g_0$ by maximizing the above quasi-likelihood function,
 \begin{align*}
 	\hat{\theta^g} = \arg \max_{\theta^g \in \Theta} \hat{L}_{n,m}(\theta^g),
 \end{align*}
 where $\Theta$ is the parameter space of $\theta^g$.
 To establish its asymptotic properties, we need the following technical conditions.

\begin{remark}
Even if the effect of the initial value is negligible,  for the finite sample, the random variable $\log RV_1$ happens to be far from the true initial value.
To handle this practical issue, we can assume that the initial value is a long-term average.
Under this condition, we can use the theoretical average value $\frac{\omega^g}{1-\gamma- \beta^g}$ as the initial value. 
That is, we additionally assume that the initial value is $H_1 (\theta^g_0)= \frac{\omega_0^g}{1-\gamma_0- \beta_0^g}$. 
 With this condition, we can obtain the same asymptotic result derived in Theorem \ref{thm:asymptotic}.
\end{remark}

 \begin{assumption} \label{assumption1} ~ 
 \begin{enumerate}[label=(\alph*)] 
 \item  $\theta^g_0 \in \Theta = \{ (\omega^g, \gamma, \beta^g); \omega_l < |\omega^g| < \omega_u,  \gamma _l  < |\gamma| < \gamma_u<1 ,   \beta_l  < |\beta| < \beta_u <1, |\gamma + \beta^g| <1  \}$, where $\omega_l, \omega_u, \gamma_l, \gamma_u, \beta_l, \beta_u$ are some known constants. 
 
 \item \label{ass:RV} $ \sup_i \bbE  \[ \left|RV_{i} - \int_{i-1}^{i  }\sigma_t^2(\theta^g_0)dt\right|^4 \] ^{1/4}   \le C m^{-1/4}$ and  $\sup_i \bbE  \[ \left| \log RV_{i} - \log \int_{i-1}^{i  }\sigma_t^2(\theta^g_0)dt\right|^4 \] ^{1/4}   \le C m^{-1/4}$


 \item $\bbE \[  \left| \int_{i-1}^i \sigma_t^2(\theta^g_0) dt\right|^4 \] \leq  C$, $\bbE \[  \sup_{\theta^g \in \Theta}   \exp( 4| H_i (\theta^g)  | )   \] \leq  C$, $\bbE \[  \sup_{\theta^g \in \Theta}   \exp( 4| \hat{H}_i (\theta^g)  | )   \] \leq  C$, and $E \[M_i ^4 \]  \leq C$ for all $i$. 

 \end{enumerate}
 \end{assumption}

\begin{remark}
Under Assumption \ref{assumption1}(a), unlike the linear GARCH-It\^o models \citep{kim2016unified, song2021volatility}, we allow the parameters to be negative. 
The condition $|\gamma + \beta^g| <1$ provides stationary properties of conditional volatilities. 
There exist  realized volatility estimators satisfying Assumption \ref{assumption1}(b) under some finite moment condition (see \citet{kim2016asymptotic, tao2011large}). 
The sufficient condition for Assumption \ref{assumption1}(c) is   that $E \[ \exp( s | \log RV_i| )  \] \leq C$ and $E \[ \exp( s | \log \int_{i-1}^{i  }\sigma_t^2(\theta^g_0)dt| )  \] \leq C$  for  $s \geq 4 / (1-\gamma_u)$. 
\end{remark}

In the following theorem, we establish the asymptotic properties of the proposed quasi-maximum likelihood estimator (QMLE).
 \begin{theorem} \label{thm:asymptotic}
Under Assumption \ref{assumption1}, we have
$$
\| \hat{\theta^g}- \theta^g_0 \|_{\max} =  O_p (n^{-1/2} +m^{-1/4}).
$$
Furthermore, we suppose that $n  m^{-1/2} \to 0$ and Assumption \ref{assumption1} is satisfied.
Then, we have
\begin{equation*}
	\sqrt{n} ( \hat{\theta^g} - \theta^g_0) \overset{d}{\to} N (0,  A V^{-1} ),
\end{equation*} 
where $A= \bbE \[ ( 1 - M_i )^2  \] $ and 
$V= \bbE \[  \frac{\partial   H_i (\theta^g) }{\partial \theta^g }  \frac{\partial    H_i (\theta^g) }{ \partial (\theta^g) ^{\top}} \big | _{\theta^g =\theta^g_0} \] $. 
 \end{theorem}

\begin{remark}
Theorem \ref{thm:asymptotic} shows that the QMLE $\hat{\theta^g}$ has the convergence rate $n^{-1/2} + m^{-1/4}$. 
The $n^{-1/2}$ term is the usual convergence rate due to the low-frequency errors, $M_i-1$. 
The $m^{-1/4}$ term is the cost to estimate the integrated volatility, which is known as the optimal rate with the presence of the micro-structure noises. 
Specifically,  by Theorem \ref{thm-integratedVol}, we have the following relationship:
 $$
  \int_{n-1} ^{n}  \sigma_{t}^2(\theta^g)  dt = \exp \left( H_n(\theta^g)   \)  M_n \text{ a.s.},
 $$
and additionally, due to the estimation error of the latent integrated volatility, we have
 $$
  RV_n = \int_{n-1} ^{n}  \sigma_{t}^2(\theta^g)  dt + E_n = \exp \left( H_n(\theta^g)   \)  M_n+ E_n \text{ a.s.}, 
 $$
where $E_n$ is the estimation error of the latent integrated volatility. 
 The error rate of $E_n$ is $m^{-1/4}$, and its asymptotic variance is specified in the literature of estimating integrated volatility  \citep{ait2010high,  barndorff2008designing, jacod2009microstructure, xiu2010quasi,  zhang2006efficient}. 
Then, the asymptotic variance of $\hat{\theta^g}$ in Theorem \ref{thm:asymptotic} has an additional term that is a function of variance of $E_n$.
For example, we have 
$$
	\| \hat{\theta^g} -\theta^g_0 \|_{\max}  \approx  C_1 \frac{\sqrt{A}} {n ^{1/2}}   +C_2 \frac{\sqrt{Avar_{RV}}} {m ^{1/4}} ,  
$$
 where $Avar_{RV}$ is the asymptotic variance of $RV$, and $C_1$ and $C_2$ are  functions of $H_i(\theta^g_0)$. 
\end{remark}

\begin{remark}
The condition $n  m^{-1/2} \to 0$ is required to remove the effect from the estimation error of the realized  volatility  when establishing the asymptotic normality. 
 However,  as in the realized GARCH model \citep{hansen2012realized}, if we assume that the conditional volatility is a function of the realized volatility estimator $RV_i$, this assumption is not required. 
\end{remark}

\subsection{Hypothesis tests}\label{test}

 In financial practices, we are interested in  statistical inferences about the GARCH parameters $(\omega^g, \gamma, \beta^g)$, such as hypothesis tests.
 In this section, we discuss how to conduct hypothesis tests for the GARCH parameters.

 Theorem \ref{thm:asymptotic} implies that 
    \begin{equation*} 
 	\sqrt{n}  ( \hat{\theta^g} - \theta^g _0 ) \overset{d}{\to}   N (0, A V^{-1} )  ,
 \end{equation*}
 where $A= \bbE \[ ( 1 - M_i )^2  \] $ and 
$V= \bbE \[  \frac{\partial   H_i (\theta^g) }{\partial \theta^g }  \frac{\partial    H_i (\theta^g) }{ \partial (\theta^g) ^{\top}} \big | _{\theta^g =\theta^g_0} \] $. 
     To evaluate the asymptotic variances of the GARCH parameter estimators, we first need to estimate $A$ and $V$. 
We use the following estimators,
$$
  \hat{A}   =  \frac{1}{n} \sum_{i=1}^n \( \frac{RV_i - \hat{H}_i (\hat{\theta^g}) }{\hat{H}_i (\hat{\theta^g})}  \)^2  \quad \text{and} \quad \hat{V}(\theta^g) = \frac{1}{n} \sum_{i=1}^n \frac{\partial \hat{H}_{i} (\theta^g )}{\partial \theta^g }\frac{\partial 
\hat{H}_{i} (\theta^g)}{\partial (\theta^g) ^{\top}},
$$
where   $\hat{H}_i (\theta^g )$ is defined in  \eqref{eq-Hhat}. 
  Under some stationary condition, we can establish their consistency.
Then, by the Slutsky's theorem, we can obtain 
\begin{equation*}
	T_{i,n}= \frac{\sqrt{n} (\hat{\theta^g}_i - \theta^g_{0 i} )} {\sqrt{\hat{A} \hat{V}_{ii}^{-1} (\hat{\theta^g}) } } \overset{d} {\to} N (0,1),
\end{equation*}
where $\hat{\theta^g}_i$  and $\theta^g_{0 i}$  are the $i$th elements of $\hat{\theta^g}$ and $\theta^g_0$, respectively, and  $\hat{V}_{ii}^{-1} (\hat{\theta^g}) $ is the $i$th diagonal element of  $\hat{V}^{-1} (\hat{\theta^g})$. 
Thus, using the proposed Z-statistics $T_{i,n}$, we can conduct the hypothesis tests based on the standard normal distribution.

\section{A simulation study} \label{SEC-simulation}
We conducted Monte-Carlo simulations to check the finite sample performance of the ERGI model.
The log-prices were generated from the ERGI model given in Definition \ref{Def-1} for $n$ days with $m$ high-frequency observations.
The model parameters were set to be $(\omega_{0}, \gamma_0,  \beta_0, \nu_{0})=(-0.1, 0.3,0.5,2)$ and $\mu_t =0$.
Then, the GARCH parameters $(\omega_0^g, \gamma_0, \beta_0 ^g)=( 0.3207, 0.3, 0.4405)$.
For the jump part, we set the intensity $\lambda_t =10$ and the jump size $|J_t|= 0.05$. 
The signs of the jump size were randomly generated. 
Let $t_{d,j}=d-1+j/m$ for $d=1,\ldots,n$ and $j=0,\ldots,m$.
We generated the noisy observations as follows:
\begin{align*}
	Y_{t_{d,j}} = X_{t_{d,j}} + \epsilon_{t_{d,j}},\qquad\text{for }d=1,\ldots,n\text{ and } j=0,\ldots, m,
\end{align*}
where $\epsilon_{t_{d,j}}$'s are i.i.d.\ normal random variables with mean zero and standard deviation $0.01 \sqrt{\int_{d-1}^d \sigma_t^2(\theta^g) dt}$.
To generate the true process, we chose $m= 11700$.
We varied $n$ from 100 to 500 and $m$ from 390 to 11700, which corresponds to the number  of minutes and 2-seconds during the open-to-close period, respectively.
We used $Y_{t_{d,j}}$ as the high-frequency observations.
To estimate the integrated volatilities, we used the jump robust pre-averaging method  \citep{ait2016increased, jacod2009microstructure} as follows:
 \begin{equation*}  
RV_d  =\frac{1} {  \psi K} \sum_{k=1}^{m-K+1} \left \{   \bar{ Y} ^2 (t_{d,k})   - \frac{1}{2} \,  \hat{ Y} ^2 (t_{d,k})    \right \}  \1_{ \{ | \bar{Y}  (t_{d,k}) |  \leq \tau_m \} },
\end{equation*}
where we take the weight function $g(x) = x \wedge (1-x)$, the bandwidth size $K= \lfloor m^{1/2} \rfloor$,
\begin{eqnarray*}
&& \bar{Y}  (t_{d,k}) = \sum_{l=1}^{K-1} g \( \frac{l}{K}\) \(   Y _{t_{d, k+l}}  - Y_{t_{d, k+l-1}} \), \quad  \psi =\int_{0}^{1} g(t)^2 dt, \cr
&&  \hat{ Y} ^2 (t_{d,k})  =    \sum_{l=1}^{K} \left \{ g\(\frac{l}{K}\)-g\(\frac{l-1}{K}\) \right \}^2 \( Y_{t_{d, k+l-1}}-Y_{t_{d,k+l-2}} \)^2,   
\end{eqnarray*}
 $\1_{\{ \cdot\} }$ is an indicator function,  and $\tau_m= c_\tau m^{- 0.235}$ is a truncation level for the constant  $c_\tau$.
 We chose $c_\tau$ as  four times the sample standard deviation of the pre-averaged prices $ m^{1/4} \bar{Y}  (t_{d,k})$. 
We estimated the parameters using the procedure in Section \ref{SEC-EST}.
We repeated the whole procedure 500 times.

To check the performance of the realized volatility estimator, we calculated squared relative errors as follows:
$$
\frac{1}{n} \sum_{i=1}^n \(\frac{RV_i - \int_{i-1}^i \sigma_t^2 (\theta_0)dt}{RV_i } \)^2.
$$
Then, we calculated the sample average of  squared relative errors over 500 simulations. 
We have the average errors 0.0117, 0.0463, and  0.10751  for $m=11700, 1170,$ and $390$, respectively.
As $m$ increases, the average errors decreases. 
This result supports the theoretical findings in the realized volatility estimator literature  \citep{ait2016increased, jacod2009microstructure}.

\begin{table}[htbp]
  \centering
  \caption{MSEs for the parameter estimates with $n=100,\,200,\,500$ and $m=390,\,1170,\, 11700$.}
    \begin{tabular}{rrccc}
 \hline
$n$	&	$m$	&	$\omega ^g$	&	$\gamma$	&	$\beta ^g$	\\ \hline
100	&	390	&	0.0854 	&	0.1312 	&	0.0468 	\\
	&	1170	&	0.0852 	&	0.1217 	&	0.0415 	\\
	&	11700	&	0.0865 	&	0.1204 	&	0.0408 	\\
	&		&		&		&		\\
200	&	390	&	0.0435 	&	0.0714 	&	0.0309 	\\
	&	1170	&	0.0428 	&	0.0690 	&	0.0280 	\\
	&	11700	&	0.0453 	&	0.0720 	&	0.0274 	\\
	&		&		&		&		\\
500	&	390	&	0.0296 	&	0.0484 	&	0.0213 	\\
	&	1170	&	0.0249 	&	0.0395 	&	0.0177 	\\
	&	11700	&	0.0244 	&	0.0367 	&	0.0171 	\\   \hline
 
    \end{tabular}%
  \label{tab:addlabel}%
\end{table}%

 Table \ref{tab:addlabel} reports the mean squared errors (MSE) of the parameter estimates $\hat{\theta^g}$ with $n=100,\,200,\,500$ and $m=390,\,1170, 11700$.
In  Table \ref{tab:addlabel}, MSEs usually decrease as the number of high-frequency observations or daily observations increases.
 This result supports the theoretical findings in Section \ref{SEC-EST}.

\begin{figure}[!ht]
  \centering
    \includegraphics[width=1\textwidth]{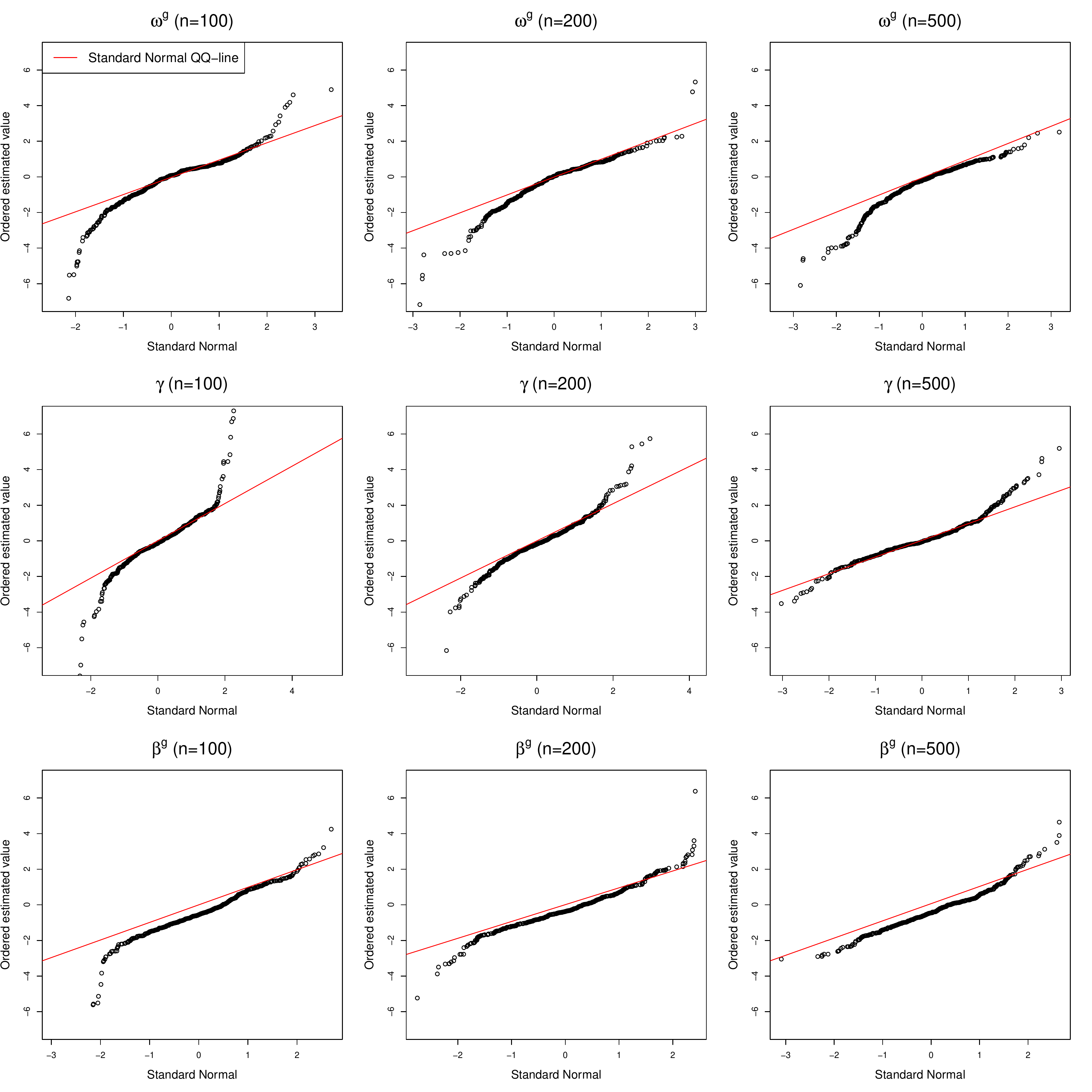}
     \caption{Standard normal QQ-plots of   the Z-statistics estimates of $\omega^g$, $\gamma$, and $\beta^g$ for $m=   390$ and  $n=100, 200, 500$.
     The red real line denotes the best linear fit line, which illustrates perfect standard normal distribution.}
     \label{Figure-QQ}
\end{figure}

To check the asymptotic normality of the GARCH parameters $(\omega^g,\gamma, \beta^g)$, we calculated the Z-statistics defined in Section \ref{test}.
In Figure \ref{Figure-QQ}, we draw the standard normal quantile-quantile plots (QQ-plots) of the Z-statistics estimates of $\omega^g$, $\gamma$, and $\beta^g$ for $m= 390$ and  $n=100, 200, 500$.
In Figure \ref{Figure-QQ}, we find that  the Z-statistics become close to  the standard normal distribution as the sample period increases.
This result supports the theoretical findings in Section \ref{SEC-EST}.
Thus, based on the proposed Z-statistics, we can conduct hypothesis tests using  the standard normal distribution.

  \begin{figure}[!ht]
  \centering
    \includegraphics[width=1\textwidth]{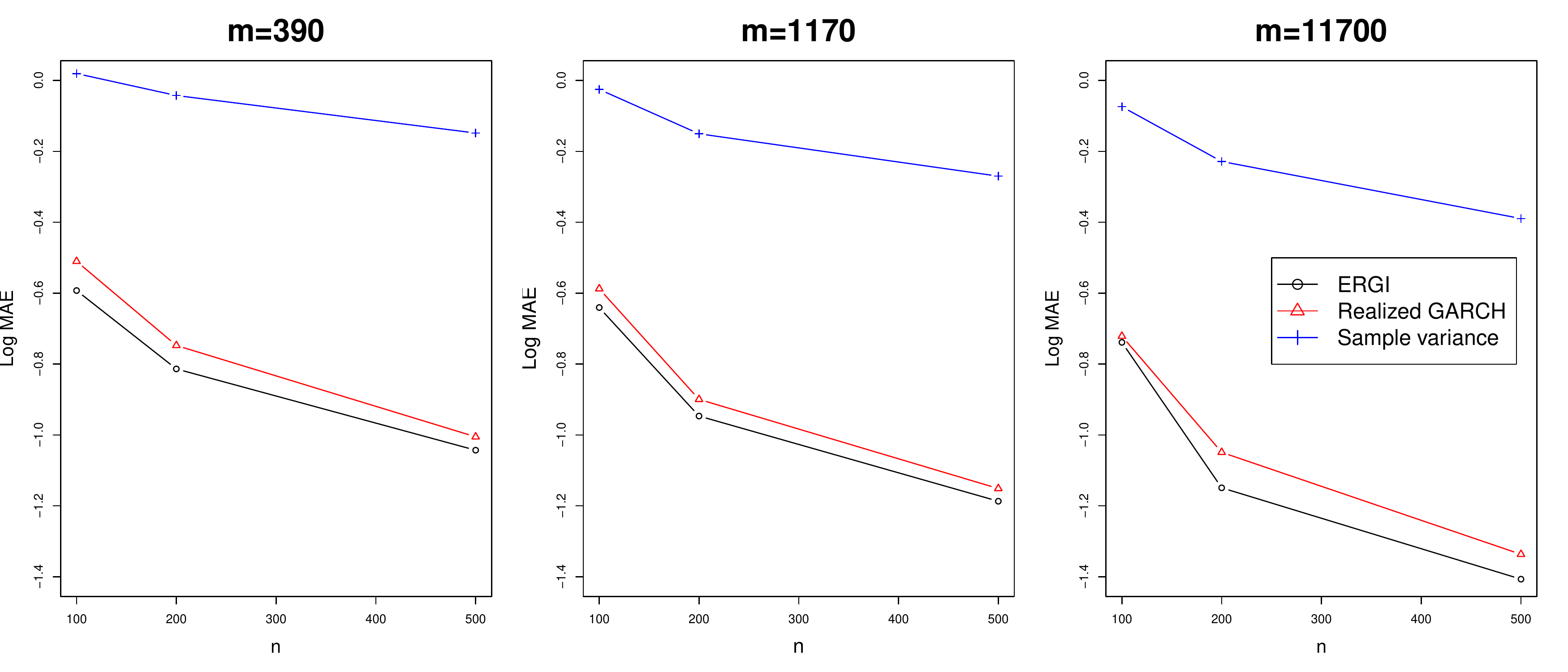}
     \caption{Log MSEs for the ERGI, realized GARCH, and PRV with $n=100, 200, 500$, and $m= 390, 1170, 11700$. }
     \label{Figure-1}
\end{figure}

 We examined the out-of-sample performance of estimating the one-day-ahead GARCH volatility $\exp ( H_{n+1} (\theta^g_0) )$. 
To estimate future GARCH volatility, we employed the proposed conditional ERGI estimator $\exp(\hat{H}_{n+1} (\hat{\theta^g}))$, realized GARCH volatility estimator \citep{hansen2012realized, song2021volatility}, and PRV of the previous day.
 For example, the realized GARCH volatility estimator is estimated based on the following  conditional volatility, 
 $$
 h_n (\theta^g) = \omega + \gamma h_{n-1} (\theta^g) + \beta RV_{n-1}.
 $$
That is, the realized GARCH volatility estimator has the usual linear GARCH structure with the realized volatilities.  
 We measured the mean squared errors with the one-day-ahead sample period over 500 samples as follows:
 $$
 \frac{1}{500} \sum_{i=1}^{500} \[ \hat{\var}_{n+1,i} - \exp\( H_{n+1,i} (\theta^g_0) \)   \]^2,
 $$
 where $ \hat{\var}_{n+1,i}$ is one of the above future volatility estimators at the $i$th sample path given the available information at time $n$. 
  Figure \ref{Figure-1} depicts the mean squared errors for the ERGI, realized GARCH, and PRV against varying the numbers of low- and high-frequency observations,   $n$ and $m$.
 In Figure  \ref{Figure-1}, we find that  the ERGI models show the best performance.
  The interesting finding is that the realized GARCH model can also capture some volatility dynamics.
This may be because even if the volatility dynamic structure is non-linear, it could have some linear dynamics. 
Especially, when the log-volatility quantities are small, by Taylor's expansion, the linear model can capture some non-linear dynamics.    
However, using the only linear structure, it cannot fully explain the non-linear dynamic structure. 
  From these results, we can conjecture that modeling appropriate dynamic structure helps account for market dynamics.

\section{Empirical study}\label{SEC-empiric}

We applied the proposed ERGI model to real trading high-frequency data. 
We obtained the top 50 trading volume assets intra-day data from January 2010 to December 2016 from the TAQ database in the Wharton Research Data Services (WRDS) system.  
We used the log-prices and employed the jump robust PRV estimation procedure defined in Section \ref{SEC-simulation} to estimate open-to-close integrated volatility.
In the empirical study, we chose the tuning parameter $c_\tau$ as  10 times the sample standard deviation of pre-averaged prices $ m^{1/4} \bar{Y}  (t_{d,k})$.
To check the accuracy of the PRV estimator, we calculated standard errors (SE) as follows. 
We first calculated  the asymptotic variance,  proposed by \citet{ait2016increased}, and divided the square root of the asymptotic variance estimator by the square root of the number of high-frequency observations.
We report data summary in Table \ref{Table-Summary}.
The number of high-frequency data is from 16,000 to 90,000 on average, and we find that the proportion of the jump variation is about  8\%  to 40\%  of the total variation on average. 
The standard error is less than 10\% of realized volatility.

\begin{table}[h]
	\centering
 
\caption{Averages of the number of high-frequency  observations, realized volatility (RV), standard error of the realized volatility estimator (SE), and jump variation (JV).}%
\scalebox{0.85}{
\begin{tabular} {ccccccccccc}
\hline
Stock	&	$\#$ of obs	&	RV$\times 10^4$	&	SE$\times 10^4$	&	JV$\times 10^4$	&&	Stock	&	$\#$ of obs	&	RV$\times 10^4$	&	SE$\times 10^4$	&	JV$\times 10^4$	\\ \hline
PG	&	16912.4 	&	0.5637 	&	0.0414 	&	0.4000 	&&	MO	&	23667.5 	&	0.6886 	&	0.0075 	&	0.2074 	\\
HBAN	&	15941.4 	&	2.7678 	&	0.0183 	&	0.3550 	&&	QCOM	&	40986.1 	&	1.3487 	&	0.0059 	&	0.2306 	\\
FCX	&	37196.1 	&	5.4242 	&	0.0222 	&	0.8249 	&&	MRK	&	36710.1 	&	0.9756 	&	0.0198 	&	0.2643 	\\
MRO	&	31241.5 	&	3.8938 	&	0.0203 	&	0.6569 	&&	GILD	&	43938.5 	&	1.8619 	&	0.0132 	&	0.4030 	\\
ORCL	&	44489.4 	&	1.3062 	&	0.0062 	&	0.1915 	&&	DAL	&	35653.1 	&	4.2791 	&	0.0219 	&	0.6506 	\\
AMD	&	22373.9 	&	5.9256 	&	0.0345 	&	0.9735 	&&	LUV	&	23756.9 	&	2.3442 	&	0.0120 	&	0.6057 	\\
AMAT	&	33556.6 	&	2.0554 	&	0.0079 	&	0.3043 	&&	T	&	38606.7 	&	0.7138 	&	0.0037 	&	0.1288 	\\
XRX	&	18877.3 	&	2.3137 	&	0.0172 	&	1.4115 	&&	CSCO	&	40328.4 	&	1.2632 	&	0.0079 	&	0.2128 	\\
WFC	&	44023.8 	&	1.4846 	&	0.0073 	&	0.3143 	&&	DIS	&	16682.3 	&	1.0435 	&	0.0108 	&	0.1842 	\\
NFLX	&	30613.5 	&	5.4658 	&	0.0293 	&	0.5557 	&&	NVDA	&	27743.1 	&	3.2549 	&	0.0151 	&	0.4210 	\\
F	&	34610.1 	&	2.2033 	&	0.0177 	&	0.4130 	&&	SLB	&	31761.7 	&	2.0279 	&	0.0094 	&	0.4409 	\\
GE	&	46327.4 	&	1.2444 	&	0.0153 	&	0.2523 	&&	BMY	&	27565.6 	&	1.1346 	&	0.0157 	&	0.2655 	\\
INTC	&	45515.7 	&	1.4031 	&	0.0076 	&	0.2464 	&&	ATVI	&	25432.0 	&	2.0102 	&	0.0100 	&	0.3804 	\\
XOM	&	45802.3 	&	0.9642 	&	0.0072 	&	0.1490 	&&	MU	&	38907.5 	&	5.1948 	&	0.0301 	&	0.9009 	\\
RF	&	19662.3 	&	3.7024 	&	0.0236 	&	0.4208 	&&	JPM	&	40898.6 	&	1.6448 	&	0.0729 	&	0.4199 	\\
DOW	&	28093.0 	&	1.9877 	&	0.0190 	&	0.3006 	&&	CVX	&	32592.7 	&	1.1472 	&	0.0067 	&	0.2205 	\\
NEM	&	29263.8 	&	3.3862 	&	0.0118 	&	0.4225 	&&	MSFT	&	61219.5 	&	1.2213 	&	0.0035 	&	0.1855 	\\
CSX	&	16106.8 	&	1.6852 	&	0.0153 	&	0.4427 	&&	BAC	&	63492.7 	&	2.4282 	&	0.0098 	&	0.3718 	\\
TXN	&	25727.4 	&	1.3822 	&	0.0072 	&	0.3163 	&&	WMT	&	30549.0 	&	0.6412 	&	0.0033 	&	0.1288 	\\
JNJ	&	31877.1 	&	0.5137 	&	0.0062 	&	0.1689 	&&	WMB	&	26970.9 	&	3.8400 	&	0.0394 	&	0.8323 	\\
VZ	&	32309.9 	&	0.7749 	&	0.0185 	&	0.2155 	&&	AAPL	&	90386.9 	&	1.4419 	&	0.0118 	&	0.3017 	\\
HST	&	21239.2 	&	2.2687 	&	0.0135 	&	0.2591 	&&	BSX	&	24163.9 	&	2.3119 	&	0.0145 	&	0.4101 	\\
MGM	&	16642.6 	&	4.5845 	&	0.0309 	&	0.6648 	&&	PFE	&	43600.5 	&	1.0821 	&	0.0160 	&	0.2296 	\\
KO	&	37024.6 	&	0.6041 	&	0.0034 	&	0.1091 	&&	HAL	&	39739.1 	&	3.2834 	&	0.0136 	&	0.4913 	\\
SCHW	&	27159.3 	&	2.0437 	&	0.1963 	&	0.3522 	&&	GLW	&	25151.0 	&	1.8426 	&	0.0155 	&	0.3186 	\\ \hline

\end{tabular}
}
\label{Table-Summary}
\end{table}

We first estimated the model parameters using the recent 1000 days data. 
From the estimated model parameters, we obtained the following conditional expected volatility for each asset 
 \begin{equation*}
 	\exp (\hat{H}_{n+1} (\hat{\theta^g}) ) \quad \text{and} \quad \hat{H}_{n+1} (\hat{\theta^g}) = \hat{\omega}^g +  \hat{\gamma} \hat{H}_{n} (\hat{\theta^g}) +\hat{\alpha}^g  \log ( RV_n ).
 \end{equation*} 
  Table \ref{Table-Estimate} reports the estimation results.    
  From Table \ref{Table-Estimate}, we show that dynamic structures can be explained by the past log-PRV,  and the coefficients of realized volatilities are statistically significant. 
 Thus, the   proposed exponential model is valid.

\begin{table}[h]
	\centering
 
\caption{ERGI model estimation results. In the parenthesis, we report the p-values. }%
\scalebox{0.85}{
\begin{tabular} {ccccccccc}
 \hline
Stock	&	$\omega^g$	&	$\gamma$	&	$\beta^g$	&&	Stock	&	$\omega^g$	&	$\gamma$	&	$\beta^g$	\\ \hline
PG	&	-1.18 (0.00)	&	0.33  (0.00)	&	0.54  (0.00)	&&	MO	&	-1.37   (0.00)	&	0.33  (0.00)	&	0.51  (0.00)	\\
HBAN	&	-1.06 (0.00)	&	0.35  (0.00)	&	0.52  (0.00)	&&	QCOM	&	-1.67  (0.00)	&	0.21  (0.00)	&	0.59  (0.00)	\\
FCX	&	-0.07  (0.00)	&	0.48 (0.00)	&	0.50  (0.00)	&&	MRK	&	-0.98  (0.00)	&	0.33  (0.00)	&	0.55  (0.00)	\\
MRO	&	-0.12  (0.00)	&	0.39  (0.00)	&	0.58  (0.00)	&&	GILD	&	-0.86   (0.00)	&	0.33 (0.00)	&	0.56  (0.00)	\\
ORCL	&	-1.60  (0.00)	&	0.18  (0.00)	&	0.64  (0.00)	&&	DAL	&	-1.91  (0.00)	&	0.18 (0.02)	&	0.57 (0.00)	\\
AMD	&	-0.50  (0.00)	&	0.50  (0.00)	&	0.42  (0.00)	&&	LUV	&	-1.45  (0.00)	&	0.33  (0.00)	&	0.49  (0.00)	\\
AMAT	&	-1.83 (0.26)	&	0.19  (0.00)	&	0.59  (0.00)	&&	T	&	-1.98 (0.01)	&	0.28 (0.00)	&	0.50  (0.00)	\\
XRX	&	-0.57  (0.00)	&	0.47 (0.00)	&	0.45  (0.00)	&&	CSCO	&	-1.83   (0.00)	&	0.18 (0.00)	&	0.61 (0.00)	\\
WFC	&	-1.17  (0.00)	&	0.24  (0.00)	&	0.62 (0.00)	&&	DIS	&	-1.41  (0.00)	&	0.23  (0.00)	&	0.61  (0.00)	\\
NFLX	&	-0.69 (0.05)	&	0.29  (0.00)	&	0.60 (0.02)	&&	NVDA	&	-1.31  (0.02)	&	0.29  (0.00)	&	0.54  (0.00)	\\
F	&	-1.56   (0.00)&	0.26 (0.03)	&	0.56  (0.00)	&&	SLB	&	-0.62   (0.00)	&	0.36  (0.00)	&	0.55  (0.00)	\\
GE	&	-1.40  (0.00)	&	0.19 (0.10)	&	0.66 (0.00)	&&	BMY	&	-1.71   (0.00)	&	0.23  (0.00)	&	0.57 (0.00)	\\
INTC	&	-1.14  (0.00)	&	0.27   (0.00)	&	0.60  (0.00)	&&	ATVI	&	-0.82   (0.00)	&	0.42  (0.00)	&	0.47  (0.00)	\\
XOM	&	-0.62 (0.00)	&	0.34  (0.00)	&	0.59  (0.00)	&&	MU	&	-1.10    (0.00)	&	0.34 (0.07)	&	0.51  (0.00)	\\
RF	&	-0.98  (0.00)	&	0.42  (0.00)	&	0.46  (0.00)	&&	JPM	&	-1.42   (0.00)	&	0.15  (0.00)	&	0.68  (0.00)	\\
DOW	&	-1.06 (0.00)	&	0.32  (0.00)	&	0.55   (0.00)	&&	CVX	&	-0.39  (0.00)	&	0.38 (0.00)	&	0.57  (0.00)	\\
NEM	&	-0.74  (0.00)&	0.35  (0.00)	&	0.55  (0.00)	&&	MSFT	&	-1.29   (0.00)	&	0.29 (0.00)	&	0.55  (0.00)	\\
CSX	&	-0.83  (0.00)	&	0.32  (0.00)	&	0.58  (0.00)	&&	BAC	&	-1.26   (0.00)	&	0.27  (0.00)	&	0.58  (0.00)	\\
TXN	&	-1.31  (0.00)	&	0.23  (0.00)	&	0.61  (0.04)	&&	WMT	&	-0.71   (0.00)	&	0.52 (0.00)	&	0.40  (0.00)	\\
JNJ	&	-1.02 (0.01)	&	0.38  (0.00)	&	0.51   (0.00)	&&	WMB	&	-0.09  (0.00)	&	0.41 (0.00)	&	0.56  (0.00)	\\
VZ	&	-1.91  (0.00)	&	0.21  (0.00)	&	0.58 (0.00)	&&	AAPL	&	-2.01   (0.00)	&	0.08  (0.00)	&	0.68  (0.00)	\\
HST	&	-0.62 (0.10)	&	0.46  (0.00)	&	0.46  (0.00)	&&	BSX	&	-1.40   (0.00)	&	0.29  (0.00)	&	0.54  (0.00)	\\
MGM	&	-1.04  (0.00)	&	0.31  (0.00)	&	0.55   (0.00)	&&	PFE	&	-0.89   (0.00)	&	0.33 (0.00)	&	0.56  (0.00)	\\
KO	&	-1.46  (0.00)	&	0.26  (0.00)	&	0.58  (0.00)	&&	HAL	&	-0.70   (0.00)	&	0.30  (0.00)	&	0.60 (0.03)	\\
SCHW	&	-1.14  (0.00)	&	0.33  (0.00)	&	0.53   (0.00)	&&	GLW	&	-1.92  (0.00)	&	0.18  (0.00)	&	0.59  (0.00)	\\ \hline
\end{tabular}
}
\label{Table-Estimate}
\end{table}

\begin{figure}[!ht]
  \centering
    \includegraphics[width=1\textwidth]{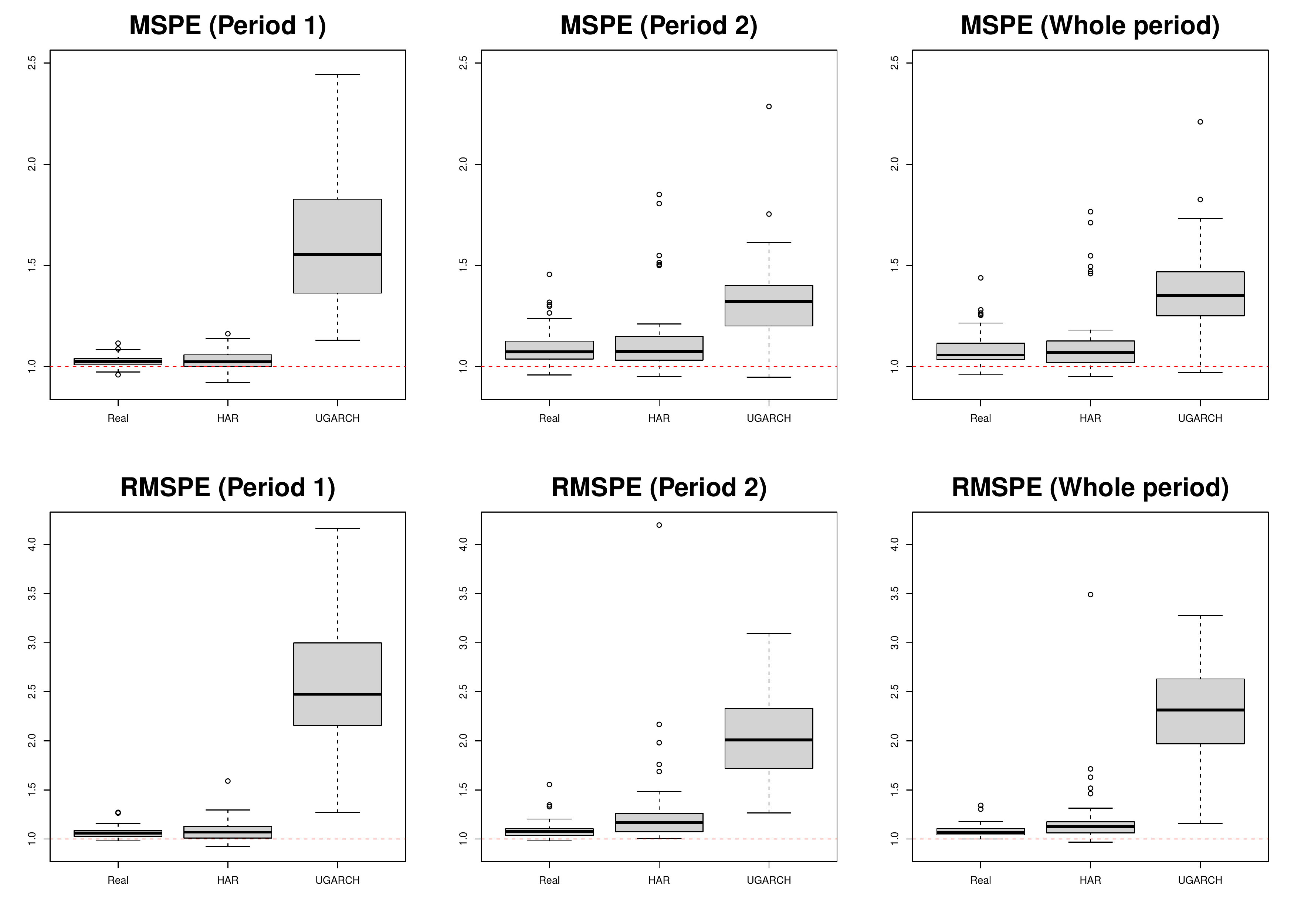}
     \caption{Box plots of relative MSPE and RMSPE for the  realized GARCH, HAR, and UGARCH with respect to the ERGI for Period 1, Period 2, and the whole period. }
     \label{Figure-box}
\end{figure}

\begin{figure}[!ht]
  \centering
    \includegraphics[width=1\textwidth]{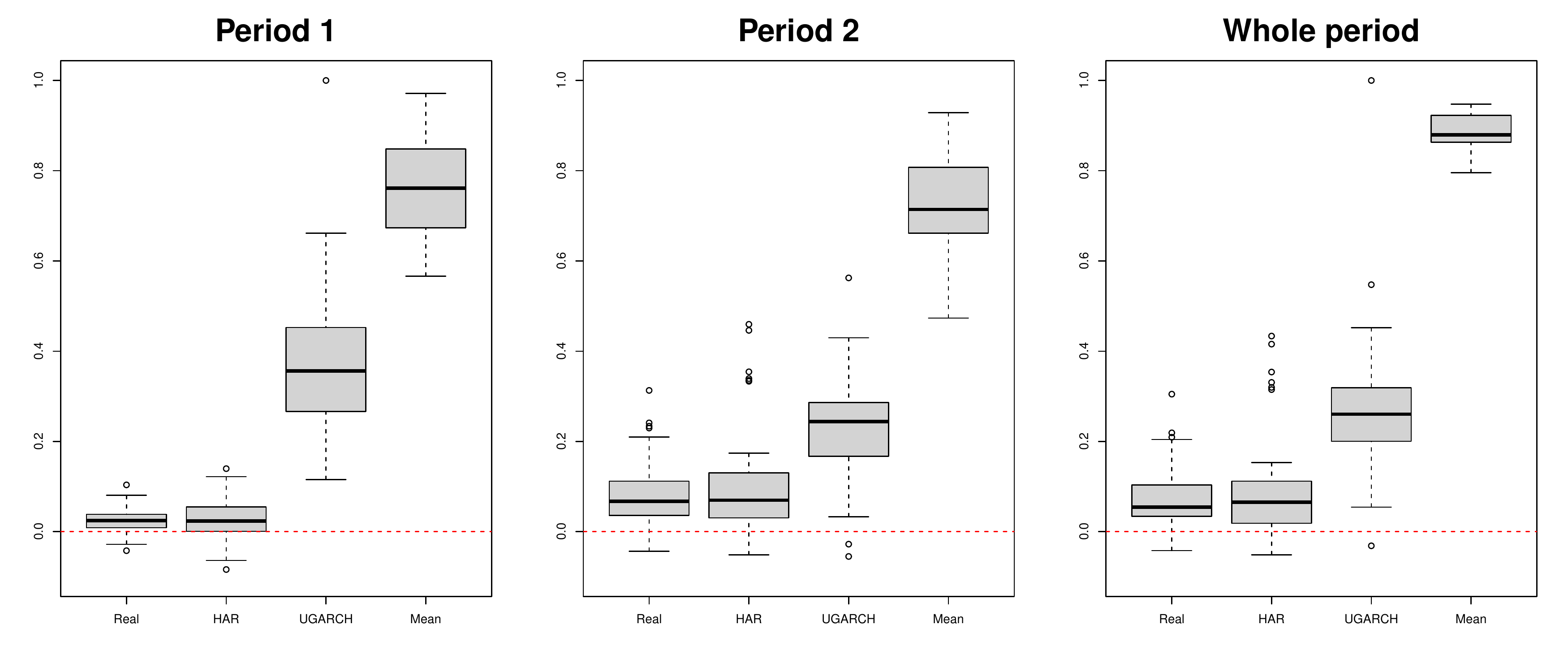}
     \caption{Box plots of OSR  for the ERGI with respect to  realized GARCH, HAR,   UGARCH, and sample mean for Period 1, Period 2, and the whole period. }
     \label{Figure-box-OSR}
\end{figure}

For comparison, we employed the realized GARCH \citep{hansen2012realized, song2021volatility}, unified GARCH-It\^o (UGARCH) \citep{kim2016unified}, and HAR \citep{corsi2009simple} models. 
To measure the performance of the volatility, we used the mean squared prediction errors (MSPE) and relative mean squared prediction errors (RMSPE)  as follows:
 \begin{eqnarray*}
  MSPE= \frac{1}{n} \sum_{i=1}^n \( Vol_i -  RV_i \)^2   \quad \text{and} \quad RMSPE=  \frac{1}{n} \sum_{i=1}^n \(\frac{ Vol_i -  RV_i}{  RV_i} \)^2,   	 
 \end{eqnarray*}
where $Vol_i $ is one of the ERGI, realized GARCH, HAR, and UGARCH. 
We used $RV_i $ as the nonparametric daily volatility estimator.
Furthermore, we calculated the out-of-sample R-square (OSR) \citep{campbell2008predicting} as follows:
 \begin{equation*}
 OSR= 1- \frac{\sum_{i=1}^n \(  RV_i - Vol_i ^* \)^2}{\sum_{i=1}^n \(  RV_i - Vol_i  \)^2},
 \end{equation*}
 where $Vol_i^*$ is  the proposed ERGI, and   $Vol_i $ is one of the  realized GARCH, HAR,  unified GARCH-It\^o, and sample mean of the in-sample $RV_i$'s. 
 We  predicted the one-day-ahead conditional expected volatility by the ERGI, realized GARCH, HAR, and UGARCH  models using the in-sample period data. 
We fixed the in-sample period as 500 days and used the rolling window scheme to estimate the parameters. 
The number of out-of-sample was 1,262.
To check the period dependency, we split the period into two equal parts. 
We denote the two periods as Period 1 and Period 2. 
Table \ref{Table-rank} reports the average rank and the number of the first rank of MSPEs and  RMSPEs  for the ERGI, realized GARCH, HAR and UGARCH for Period 1, Period 2, and the whole period  over the 50 assets.  
Figure \ref{Figure-box} depicts the relative MSPE and RMSPE for the  realized GARCH, HAR, and UGARCH with respect to the ERGI for Period 1, Period 2, and the whole period.
Figure \ref{Figure-box-OSR} draws the OSR for the ERGI with respect to  realized GARCH, HAR,   UGARCH, and sample mean for Period 1, Period 2, and the whole period.
From Table \ref{Table-rank} and Figures \ref{Figure-box}--\ref{Figure-box-OSR}, we find that the realized volatility-based model, such as the ERGI, realized GARCH, and HAR models, perform better than the UGARCH model, which incorporates the squared open-to-close returns as the innovation. 
 That is, incorporating the realized volatility helps account for the volatility dynamics. 
When comparing the realized volatility-based models, the proposed ERGI model shows the best performance.
From this result, we can conjecture that the non-linear exponential form with realized volatilities helps explain the market dynamics.

\begin{table}[h]
	\centering
 \caption{Average rank of MSPEs and RMSPEs for the ERGI,  realized GARCH, HAR and UGARCH for Period 1, Period 2, and the whole period.
In the parenthesis, we report the number of the first rank among competitors.}%
 
\scalebox{0.65}{
    \begin{tabular}{rccccc ccccc cccc}
 \hline
 & \multicolumn{4}{c}{Period 1} && \multicolumn{4}{c}{ Period 2} &&  \multicolumn{4}{c}{Whole period}\\  \cline{2-5}   \cline{7-10}   \cline{12-15} 
	&	ERGI	&	Real	&	HAR	&	UGARCH	&&	ERGI	&	Real	&	HAR	&	UGARCH	&&	ERGI	&	Real	&	HAR	&	UGARCH	\\
MSPE	&	1.4 (33)	&	2.3 (5)	&	2.2 (12)	&	4.0 (0)	&&	1.1 (42)	&	2.5 (2)	&	2.6 (4)	&	3.6 (2)	&&	1.2 (41)	&	2.5 (2)	&	2.4 (6)	&	3.7 (1)	\\
RMSPE	&	1.2 (38)	&	 2.4 (3)	&	2.3 (9)	&	4.0 (0)	&&	1.0 (48)	&	2.2 (1)	&	2.8 (0)	&	3.8 (1)	&&	1.0 (47)	&	2.3 (0)	&	2.6 (3)	&	3.9 (0)	\\ \hline

\end{tabular}
 }
\label{Table-rank}
\end{table}

\begin{figure}[!ht]
  \centering
    \includegraphics[width=1\textwidth]{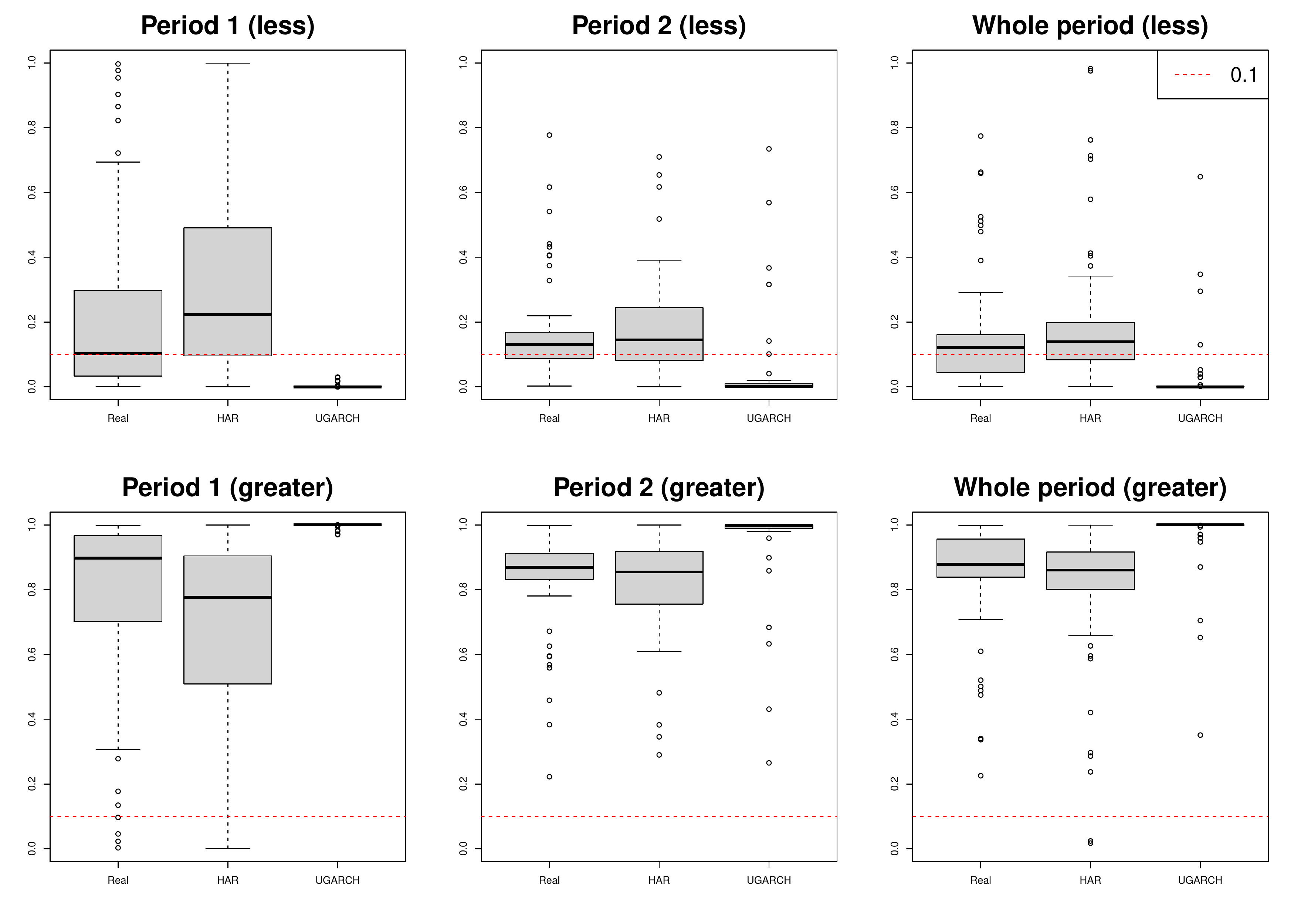}
     \caption{Box plots of p-values of the  less and greater DM tests  for the ERGI versus one of the realized GARCH, HAR, and   UGARCH  for Period 1, Period 2, and the whole period. }
     \label{Figure-box-pvalue}
\end{figure}

To further compare the predictive accuracy among  the ERGI, realized GARCH, HAR, and UGARCH  models, we conducted Diebold-Mariano tests \citep{diebold2002comparing} as follows.
We first calculated the residuals for the four models:
$$
	e_i = RV_i - Vol_i,
$$
where $Vol_i $ is one of the ERGI, realized GARCH, HAR, and UGARCH, and $RV_i$ is the non-parametric realized volatility.
We define 
$$
	d_i =  e_i^{*2} - e_i^2 ,
$$
where $e_i^*$ is the residuals from the ERGI and $e_i$ is the residuals from one of   realized GARCH, HAR, and UGARCH. 
Then, we conducted hypothesis tests for 
$$
	H_0': \mathbb{E} [ d_i ] = 0 \quad \text{v.s.}  \quad  H_1: \mathbb{E} [ d_i ] <  0 \text{ ( or  } \mathbb{E} [ d_i ] >  0 ).
$$
The first alternative statement ($\mathbb{E} [ d_i ] <  0$) is to test whether the ERGI is better, while the second alternative statement  ($\mathbb{E} [ d_i ] >  0$) is to test whether other model is better than the ERGI. 
 We call them ``less'' and ``greater'' tests, respectively. 
 Figure \ref{Figure-box-pvalue} depicts box plots of p-values of the less and greater DM tests  for the ERGI versus one of the realized GARCH, HAR, and   UGARCH  for Period 1, Period 2, and the whole period.
 From Figure \ref{Figure-box-pvalue}, the less tests show that p-values of 22, 16, and 46 assets for realized GARCH, HAR, and UGARCH models, respectively, were less than 10\% over the whole period.
 In contrast,   the greater tests indicate that a couple of assets for the HAR model have significant p-values over the whole period.
 From these results, although the ERGI does not give significant better predictive accuracy for all assets, we can conclude that for most assets, the ERGI is at least not worse than other models, and, for some assets, the ERGI shows significantly better performance than the other models.

\begin{figure}[!ht]
  \centering
    \includegraphics[width=1\textwidth]{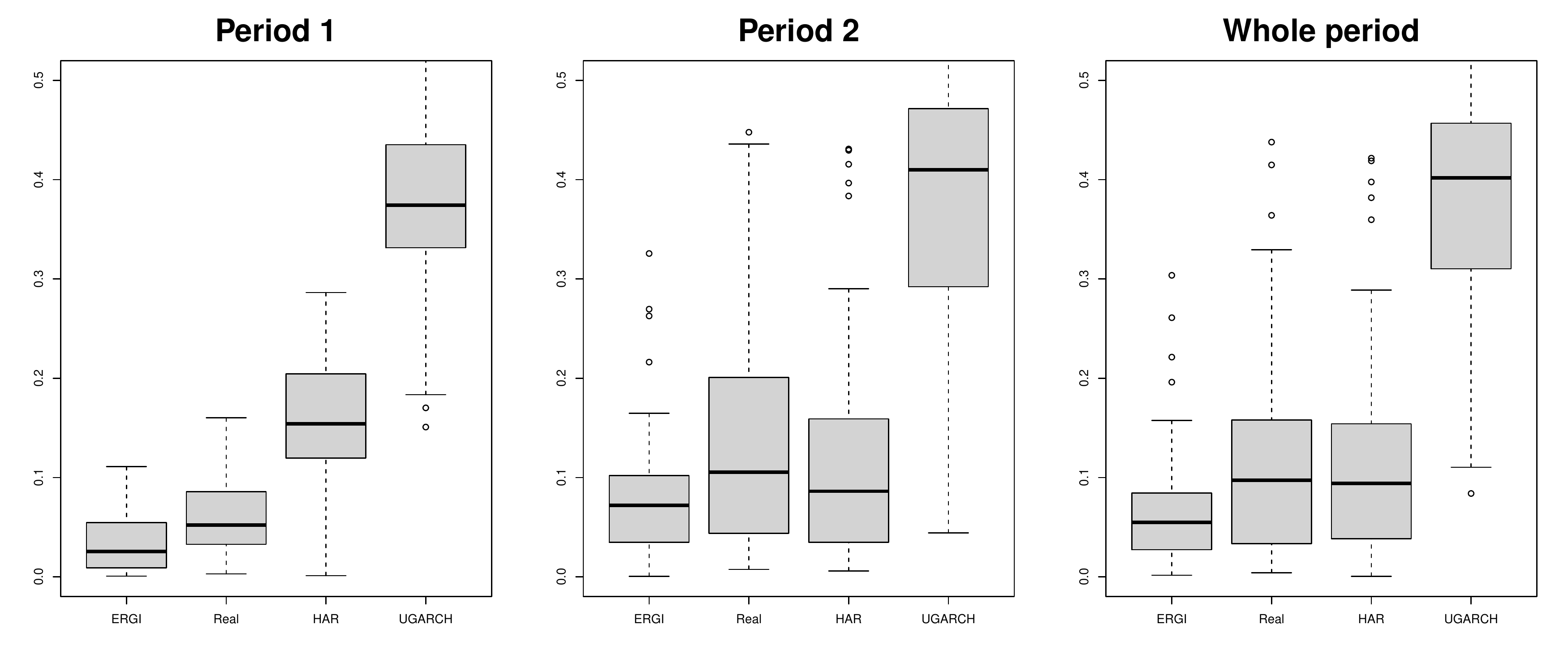}
     \caption{Box plots of the first order autocorrelation for the residuals of  the ERGI, realized GARCH, HAR and UGARCH  for Period 1, Period 2, and whole period. }
     \label{Figure-box2}
\end{figure}

To check the volatility persistence of the nonparametric volatility, we studied  the residuals between the nonparametric volatility and estimated conditional volatilities, $Vol_i -  RV_i$, where $Vol_i $ is the  predicted one-day-ahead conditional expected volatility by ERGI, realized GARCH, HAR, and UGARCH using the in-sample period data. 
Then, we checked their autocorrelations. 
Table \ref{Table-rank2} reports the average rank and  number of the first rank of the first order autocorrelation  for the ERGI, realized GARCH, HAR, and UGARCH for Period 1, Period 2, and the whole period over the 50 assets. 
Figure \ref{Figure-box2} provides the box plots of the first order autocorrelation   for the ERGI, realized GARCH, HAR, and UGARCH for Period 1, Period 2, and the whole period over the 50 assets. 
From Table \ref{Table-rank2} and Figure \ref{Figure-box2}, we find that the ERGI has relatively small autocorrelations. 
That is, the ERGI model can reduce the volatility persistence.   
These numerical results provide evidence to conclude that the non-linear exponential auto-regressive structure helps explain the market dynamics in the volatility time series.

\begin{table}[h]
	\centering
 \caption{Average rank of the first order autocorrelation for the ERGI,  realized GARCH, HAR and UGARCH for Period 1, Period 2, and the whole period. 
In the parenthesis, we report the number of the first rank among competitors.}%
 
\scalebox{0.7}{
    \begin{tabular}{ccccc ccccc cccc}
 \hline
    \multicolumn{4}{c}{Period 1} && \multicolumn{4}{c}{ Period 2} &&  \multicolumn{4}{c}{Whole period}\\  \cline{1-4}   \cline{6-9}   \cline{11-14} 
ERGI	&	Real	&	HAR	&	UGARCH	&&	ERGI	&	Real	&	HAR	&	UGARCH	&&	ERGI	&	Real	&	HAR	&	UGARCH	\\ \hline
1.2 (40)	&	1.9 (9)	&	2.9 (1)	&	3.9 (0)	&&	1.8 (18)	&	2.3 (14)	&	2.3 (13)	&	3.6 (5)	&&	1.6 (28)	&	2.2 (12)	&	2.5 (8)	&	3.6 (2)	\\ \hline
\end{tabular}
 }
\label{Table-rank2}
\end{table}

\section{Conclusions}\label{SEC-conclude}
 
In this paper, we propose a novel jump diffusion process to model the non-linear auto-regressive structure of the realized volatility.
We employ the exponential GARCH structure. 
By introducing a continuous instantaneous volatility process whose integrated volatility follows the exponential realized GARCH structure, we fill the gap between the empirical discrete-time non-linear volatility model with the realized volatility and high-frequency based continuous-time diffusion process.
That is, this paper provides rigorous mathematical background to understand the exponential realized GARCH structure. 
To estimate the model parameter, we propose the quasi-maximum likelihood estimation procedure and establish its asymptotic properties.
From the empirical study, we find the benefits of incorporating the non-linear exponential realized GARCH.

In this paper, we focus on the continuous part of log-return processes for the open-to-close period. 
However, it is important and interesting to study dynamic structures of the jump variation and close-to-open returns.
We leave this for future study.

\section{Proofs} \label{SEC-proof}

\subsection{ Proof of Theorem \ref{thm-integratedVol}}
 
  \begin{lemma}\label{exp-lemma}
Under the ERGI model in Definition \ref{Def-1}, we have for $t\in (n-1,n]$,
    \begin{align*}
 	\frac{1}{t-n+1}\int_{n-1}^t \sigma_s^2 (\theta) ds = \sigma_{n-1}^2(\theta) e^{\int_{n-1}^t b_s(\theta) ds} \text{ a.s.}\numberthis \label{eq:sde}
 \end{align*} 
 \end{lemma}
  \textbf{Proof of Lemma \ref{exp-lemma}.}
 We have
 \begin{eqnarray*}
 	d ((t-[t])\bar{\sigma}_t^2 (\theta))  &=&  (t-[t]) d \bar{\sigma}_t^2 (\theta) + \bar{\sigma}_t^2 (\theta) dt \cr
 	&=&   \sigma_t^2(\theta) dt  \cr
 		&=&  \bar{\sigma}_t^2 (\theta) \left( 1 + (t-[t])b_t (\theta) \right) dt \text{ a.s.}
 \end{eqnarray*}
 Thus, we have
 \begin{eqnarray*}
	 (t-[t]) d \bar{\sigma}_t^2 (\theta) &=&  \bar{\sigma}_t^2 (\theta) \left( 1 + (t-[t])b_t (\theta) \right) dt -\bar{\sigma}_t^2 (\theta) dt  \cr
	 	&=& (t-[t]) \bar{\sigma}_t^2 (\theta)b_t (\theta)   dt \text{ a.s.}
 \end{eqnarray*}
This implies that
 \begin{eqnarray*}
  d \bar{\sigma}_t^2 (\theta) =    \bar{\sigma}_t^2 (\theta)b_t (\theta)   dt  \text{ a.s.}
 \end{eqnarray*}
 and 
 \begin{equation*}
 	\frac{1}{t-n+1}\int_{n-1}^t \sigma_s^2 (\theta) ds = \sigma_{n-1}^2(\theta) e^{\int_{n-1}^t b_s(\theta) ds} \text{ a.s.} \qquad\text{for }t\in (n-1,n].
 \end{equation*}
\endpf 
 
 \textbf{Proof of Theorem \ref{thm-integratedVol}.}
  First, we consider $(a)$.
 By the It\^o's lemma, we have 
 \begin{eqnarray*}
 	R (k) &=& \int_{n-1} ^{n} \frac{(n-t)^k }{k!} b_t (\theta) dt  \cr
 		&=&  \frac{   \omega   }{ (k+2) !}    +   \nu \( \frac{1}{(k+2)!} - \frac{2}{(k+3)!} \) \cr
 			&&+ \left \{  \frac{ 1 }{(k+1)!} +  \frac{\gamma -1  }{(k+2)!}  \right \} b_{n-1} (\theta)  \cr
  			&& - \left\{  \frac{\beta }{ (k+1) ! } +   \frac{\beta^* - \beta  }{ (k+2)! } -\frac{2 \beta^*   }{ (k+3)! }  \right\} \log \sigma_{n-1}^2 (\theta) 	\cr
 			&& +  2 \nu  \  \int_{n }^{n -1}   \frac{ (n-t)^{k+2} }{k! (k+2)}  Z_t   dW_t   +    \beta   \int_{n-1}^{n}   \frac{(n- t)^k }{k!} \log \bar{\sigma}_t^2 (\theta) dt  \text{ a.s.},
 \end{eqnarray*}
and,  by Lemma \ref{exp-lemma},   we have 
 \begin{eqnarray*}
 &&\int_{n-1}^{n}   \frac{(n- t)^k }{k!} \log \bar{\sigma}_t^2 (\theta) dt \cr
 &&=\int_{n-1}^{n}   \frac{(n- t)^k }{k!} \left( \int_{n-1} ^{t}b_s (\theta) d s + \log \sigma_{n-1}^2 (\theta) \right)  dt \cr
 &&= \frac{1}{(k+1)!} \log \sigma_{n-1}^2 (\theta) +  \int_{n-1}^{n}  b_s (\theta)  \int_{s} ^{n}   \frac{(n- t)^k }{k!} dt ds \cr
 &&= \frac{1}{(k+1)!} \log \sigma_{n-1}^2 (\theta)+ R (k+1) \text{ a.s.}
 \end{eqnarray*}
  Thus, we have
 \begin{eqnarray*}
R (0)&=& \int_{n-1} ^{n}  b_t (\theta)  dt  \\
	& =  & \sum_{k=0}^\infty     \omega    \beta ^{-2}  \frac{ \beta    ^{k+2} }{ (k+2) !}  + \nu \( \beta^{-2}   \frac{\beta^{k+2}  }{(k+2)!}-\beta^{-3}   \frac{\beta^{k+3}  }{(k+3)!} \) \\
	&& +  \sum_{k=0}^\infty b_{n-1} (\theta)   \left \{   \beta^{-1} \frac{ \beta^{k+1}  }{(k+1)!} +(\gamma-1 ) \beta^{-2} \frac{2\beta^{k+2}  }{(k+2)!} \right \}  \\
	&& -  \sum_{k=0}^\infty \log  \sigma_{n-1}^2 (\theta)   \left \{  \left( \beta^* - \beta  \right) \beta^{-2}  \frac{\beta^{k+2}  }{(k+2)!}  -  2\beta^* \beta^{-3}   \frac{\beta^{k+3}  }{(k+3)!}  \right \} \\
	&& + \sum_{k=0}^{\infty} 2  \nu   \int_{n-1}^{n -1}   \beta   ^{k} \frac{(n -t) ^{k+2} (k+1) }{ (k+2)!}  Z_t dW_t  \\
	&=&  \omega \varrho_2 + \nu (\varrho_2 - 2 \varrho_3) - \{ (\beta^* -\beta) \varrho_2 - 2 \beta^* \varrho_3  \} \log \sigma_{n-1}^2 (\theta) \cr
		&&  +   \{    \varrho_{1} + (\gamma -1 )  \varrho_{2} \}  b_{n-1} (\theta) + D_n  \cr
		& = &  \omega \varrho_2  + \nu (\varrho_2 - 2 \varrho_3) - \log \sigma_{n-1}^2 (\theta)   + \varrho  b_{n-1} (\theta) + D_n \cr
		&=& h_n (\theta) -\log \sigma_{n-1}^2 (\theta) + D_n \text{ a.s.} 
 \end{eqnarray*} 
Then, by \eqref{eq:sde}, we have
 \begin{eqnarray*} 
  \int_{n-1} ^{n}  \sigma_t^2 (\theta)  dt &=& \sigma_{n-1}^2 \exp \(  \int_{n-1} ^{n} b_t(\theta) dt \) \cr
  	&=&  \exp \left( h_n(\theta)  + D_n \right) \text{ a.s.} 
 \end{eqnarray*}

 For $(b)$, since $\bbE [ \exp (D_n) ]$ is a constant, we have 
\begin{align*}
	\int_{n-1} ^{n}  \sigma_t^2 (\theta)  dt &= \exp \( H_n (\theta)    \) M_n,
\end{align*} 
and we obtain $(b)$.
\endpf

\subsection{Proof of Theorem \ref{thm:asymptotic}}

To simplify the notation, we use $\theta$ for the GARCH model parameters $\theta^g$.
For derivatives of any given function $f$ at $x_0$, we denote $\frac{\partial  f(x_0)}{\partial x} = \frac{\partial f(x)}{\partial x} \bigr|_{x=x_0}$.
 Define
 \begin{eqnarray*}
 &&\hat{L}_{n,m}(\theta)  = - \frac{1}{n}\sum_{i=1}^n \left \{  \hat{H}_i(\theta) + \frac{ RV_{i}   }{\exp\left( \hat{H}_i(\theta) \right)}  \right \}  \quad  \text{ and }  \quad \hat{s}_{n,m}  (\theta) = \frac{ \partial \hat{L}_{n,m} (\theta) }{\partial \theta}	; \\
 &&\hat{L}_{n}(\theta)  = - \frac{1}{n}\sum_{i=1}^n \left \{  H_i(\theta) + \frac{ \int_{i-1}^i \sigma_t^2(\theta_0) dt  }{\exp\left( H_i(\theta) \right)}   \right \}  \quad  \text{ and }  \quad \hat{s}_{n}  (\theta) = \frac{ \partial  \hat{L}_{n} (\theta) }{\partial \theta}	; 	\\
 &&L_{n}(\theta)  = - \frac{1}{n}\sum_{i=1}^n \left \{  H_i(\theta) + \frac{ \exp \left( H_i(\theta_0) \right)   }{\exp\left( H_i(\theta) \right)}   \right \}  \quad  \text{ and }  \quad s_{n}  (\theta) = \frac{  \partial  L_{n} (\theta) }{\partial \theta}.
 \end{eqnarray*} 
 Since the dependence of $H_i(\theta)$ on the initial value decays exponentially \citep{kim2016unified}, we use the true initial value $H_0(\theta_0)$ for the rest of the proofs without of loss of generality.

  \begin{lemma} \label{lemma:in_prob_para}
  Under the assumptions of Theorem \ref{thm:asymptotic},  we have $\hat{\theta} \xrightarrow{p} \theta_0$.
  \end{lemma}
  
  \textbf{ Proof of Lemma \ref{lemma:in_prob_para}.} 
  We first show the uniform convergence of $\hat{L}_{n,m}(\theta)$. 
  That is, we need to show  
 \begin{align*}
  \sup_{\theta \in \Theta}\left| \hat{L}_{n,m}(\theta) - L_n (\theta) \right| & \leq   \sup_{\theta \in \Theta}\left| \hat{L}_{n,m}(\theta) - \hat{L}_n (\theta) \right|  + \sup_{\theta \in \Theta}\left| \hat{L}_{n}(\theta) - L_n (\theta) \right|  	\\
  &= o_p(1).
\end{align*}  
For $\sup_{\theta \in \Theta}\left| \hat{L}_{n,m}(\theta) - \hat{L}_n (\theta) \right|$, we have
\begin{align*}
	&\sup_{\theta \in \Theta}\left| \hat{L}_{n,m}(\theta) - \hat{L}_n (\theta) \right|  	\\
	&\le \sup_{\theta \in \Theta} \Bigg\{\frac{1}{n}\left|\sum_{i=1}^n \hat{H}_i(\theta) - H_i(\theta) \right|  	 +  \frac{1}{n}\left|\sum_{i=1}^n \frac{RV_{i}   - \int_{i-1}^i \sigma_t^2(\theta_0) dt }{\exp\left( \hat{H}_i(\theta) \right)}\right|  	\\
	&\qquad\qquad + \frac{1}{n}\left|\sum_{i=1}^n   \int_{i-1}^i \sigma_t^2(\theta_0) dt   \left( e^{- \hat{H}_i(\theta)} - e^{- H_i(\theta) }\right) \right|  \Bigg\}	\\
	&= \text{(I)} + \text{(II)} + \text{(III)}.
\end{align*}
For (I), we have
\begin{eqnarray} \label{eq1-Lemma-2}
	\bbE[\text{(I)}] &\leq& \frac{1}{n}\sum_{i=1}^n \bbE \[ \Bigg| \sup_{\theta \in \Theta} \sum_{k=1}^{i-1}\beta\gamma^{k-1} \left( \log RV_{i-k} - \log IV_{i-k}  \right) \Bigg|\] 	\cr
	&\leq &  \frac{C}{n}\sum_{i=1}^n \sum_{k=1}^{i-1}\gamma^{k-1}_{u}   \bbE  \[   |\log RV_{i-k} -\log IV_{i-k} | \]    	\cr
	&\leq& C m^{-1/4},
\end{eqnarray}
where $IV_{i} =  \int_{i-1}^i \sigma_t^2 (\theta_0) dt$, and the last inequality is due to Assumption  \ref{assumption1}\ref{ass:RV}. 
Consider (II). 
 By Assumption  \ref{assumption1}\ref{ass:RV} and (c), we have
\begin{align*}
	\bbE[\text{(II)}] &\le\frac{C}{n}\sum_{i=1}^n \bbE  \[ \left|  RV_i - IV_i \right| ^2  \] \bbE  \[ \sup_{\theta \in \Theta}  \left|  \exp\left( | \hat{H}_i(\theta)| \right)  \right| ^2  \] 	 \le C m^{-1/4}.  
\end{align*}
For (III), we have
\begin{eqnarray*}
	\bbE[\text{(III)}] & \leq & \frac{C}{n} \sum_{i=1}^n \bbE  \[ \sup_{\theta \in \Theta}  \left(e^{- \hat{H}_i(\theta)} - e^{- H_i(\theta)} \right)  ^2 \]^{1/2}  \cr
		&\leq& \frac{C}{n} \sum_{i=1}^n  \bbE  \[ \sup_{\theta \in \Theta} \left (e^{4 | \hat{H}_i(\theta)|}  +  e^{ 4|H_i(\theta)|} \right)   \]^{1/4} \bbE  \[ \sup_{\theta \in \Theta}  \left ( \hat{H}_i(\theta) -  H_i(\theta)  \right)  ^4 \]^{1/4}  \cr 
	&\leq& C m^{-1/4}, 
\end{eqnarray*}
where the first and second inequalities are due to Holder's inequality and Taylor's expansion, respectively, and the last inequality can be showed similar to the proof of \eqref{eq1-Lemma-2} with Assumption  \ref{assumption1}(c).
Thus, we have
\begin{equation}\label{r1-Lemma-2}
	\sup_{\theta \in \Theta}\left| \hat{L}_{n,m}(\theta) - \hat{L}_n (\theta) \right| = o_p(1). 
\end{equation}

We consider $\sup_{\theta \in \Theta}\left| \hat{L}_{n}(\theta) - L_n (\theta) \right|$.
We have
\begin{eqnarray*}
	\hat{L}_{n}(\theta) - L_n (\theta)  =- \frac{1}{n} \sum_{i=1}^n \frac{e^{H_i (\theta_0) }}{e^{H_i(\theta)} }  ( M_i -1), 
\end{eqnarray*}
which is a martingale process for any given $\theta$. 
Thus, by martingale convergence theorem, we can show its pointwise convergence. 
To show its uniform convergence, we need to show the stochastic continuity for $G_n (\theta)=\hat{L}_{n}(\theta) - L_n (\theta) $.
By the Taylor's expansion and the mean value theorem, there exits $\theta^*$ between $\theta$ and $\theta^{\prime}$ such that
\begin{eqnarray*}
	|G_n (\theta) - G_n (\theta^{\prime}) | &=&  \left |  \frac{1}{n} \sum_{i=1}^n \frac{e^{H_i (\theta_0) }}{e^{H_i(\theta^*)} } \frac{\partial H_i (\theta^*)}{ \partial \theta}   ( M_i -1)  ( \theta - \theta ^{\prime} ) \right |  \cr
		&\leq &C \frac{1}{n} \sum_{i=1}^n   \sup_{\theta^* \in \Theta }  \left \| \frac{e^{H_i (\theta_0) }}{e^{H_i(\theta^*)} } \frac{\partial H_i (\theta^*)}{ \partial \theta}   ( M_i -1) \right \|_{\max} \| \theta - \theta^{\prime} \|_{\max}. 
\end{eqnarray*}
By Assumption \ref{assumption1}(c), we have $\bbE\[e^{ 4|H_i(\theta)| }\] \leq C $.
Then, similar to the proofs of Lemma 3 in  \citet{kim2016unified},  we can show 
$$
\frac{1}{n} \sum_{i=1}^n   \sup_{\theta^* \in \Theta }  \left \| \frac{e^{H_i (\theta_0) }}{e^{H_i(\theta^*)} } \frac{\partial H_i (\theta^*)}{ \partial \theta}   ( M_i -1) \right \|_{\max} =O_p(1).
$$ 
Thus, $G_n(\theta)$ satisfies the weak Lipschitz condition, so, by Theorem 4 in \citet{andrews1992generic}, we can show the uniform convergence. 
Therefore, we have
\begin{equation}\label{r2-Lemma-2}
	\sup_{\theta \in \Theta}\left| \hat{L}_{n}(\theta) - L_n (\theta) \right| = o_p(1). 
\end{equation}
By \eqref{r1-Lemma-2} and  \eqref{r2-Lemma-2}, we show the uniform convergence of $\hat{L}_{n,m}(\theta)$.

When $\exp (H_i( \theta_0)) = \exp (H_i (\theta))$ for all $i$, $L_n(\theta)$ is maximized. 
Obviously, $\theta_0$ is one of the solutions. 
Suppose that there exists $\theta_*$ such that  $\exp (H_i( \theta_0)) = \exp (H_i (\theta_*))$ a.s. for all $i$.
Since the exponential function is a  strictly increasing function, we have $H_i( \theta_0) -  H_i (\theta_*)=0$ a.s. for all $i$.
Thus, $\theta_0$ and $\theta_* = (\omega_*^g, \gamma_*, \beta_*^g)$ satisfy
\begin{equation*}
\begin{pmatrix}
1 & H_1 (\theta_0)  & \log IV_1   \\ 
 1& H_2 (\theta_0)  & \log IV_2 \\ 
 \vdots& \vdots  & \vdots \\ 
 1&  H_{n-1} (\theta_0)  & \log IV_{n-1} 
\end{pmatrix}
\begin{pmatrix}
\omega_*^g - \omega_0^g   \\ 
\gamma_*- \gamma_0 \\   
 \beta_*^g - \beta_0 ^g
\end{pmatrix}
\equiv  M \begin{pmatrix}
\omega_*^g - \omega_0^g   \\ 
\gamma_*- \gamma_0 \\   
 \beta_*^g - \beta_0 ^g
\end{pmatrix}=0 \text{ a.s.}
\end{equation*}
Since $IV_i$'s are non-degenerating, $M$ is of full rank. 
Then, $M ^\top M$ is invertible, which implies $\theta_0 = \theta_*$ a.s.
Therefore, $L_n (\theta)$ has the unique maximizer $\theta_0$. 
Then, by  Theorem 1 in \citet{xiu2010quasi}, with the  uniform convergence of $\hat{L}_{n,m}(\theta)$, we can show $\hat{\theta} \overset{p}{\to} \theta_0$. 
\endpf

\textbf{Proof of Theorem  \ref{thm:asymptotic}.}
The mean value theorem and Taylor's expansion, there exists $\theta^*$  between $\hat{\theta}$ and $\theta_0$ such that
\begin{equation*}
	 \hat{s}_{n,m}(\hat{\theta}) -  \hat{s}_{n,m}(\theta_0) = -  \hat{s}_{n,m}(\theta_0) = \triangledown \hat{s}_{n,m}(\theta^*)(\hat{\theta} - \theta_0),
\end{equation*}
 where $\triangledown \hat{s}_{n,m}(\theta^*) =\frac {\partial \hat{s}_{n,m}(\theta^*)}{\partial \theta ^{\top} }$.
 We first consider $ \hat{s}_{n,m}(\theta_0)$. 
 We have
 \begin{eqnarray}\label{eq1-thm-asym}
 	- \hat{s}_{n,m}(\theta_0)  &=&  \frac{1}{n} \sum_{i=1}^n \left \{ 1 - e^{-\hat{H}_i (\theta_0) } RV_i \right \} \frac{\partial \hat{H}_i (\theta_0) }{\partial \theta} \cr
 		&=&  \frac{1}{n} \sum_{i=1}^n \left \{ 1 - e^{-H_i (\theta_0) } IV_i \right \} \frac{\partial H_i (\theta_0) }{\partial \theta} +O_p(m^{-1/4})  \cr
 		 		&=&  \frac{1}{n} \sum_{i=1}^n ( 1 - M_i ) \frac{\partial H_i (\theta_0) }{\partial \theta} +O_p(m^{-1/4})  \cr
 		 		&=& O_p ( n^{-1/2} + m^{-1/4}),
 \end{eqnarray}
 where the second equality can be showed similar to the proofs of Lemma \ref{lemma:in_prob_para}, and the last equality is due to the martingale convergence theorem.

  We consider $\triangledown \hat{s}_{n,m}(\theta^*)$. 
Similar to the proofs of \eqref{eq1-thm-asym} with the consistency of $\hat{\theta}$, we can show
 \begin{eqnarray*}
 	\triangledown \hat{s}_{n,m}(\theta^*)  &=&-  \frac{1}{n} \sum_{i=1}^n \left \{ 1 - e^{-\hat{H}_i (\theta^*) } RV_i \right \} \frac{\partial ^2  \hat{H}_i (\theta^*) }{\partial \theta \partial \theta ^{\top}}     - \frac{1}{n} \sum_{i=1}^n  e^{-\hat{H}_i (\theta^*) } RV_i  \frac{\partial  \hat{H}_i (\theta^*) }{\partial \theta  }  \frac{\partial    \hat{H}_i (\theta^*) }{ \partial \theta ^{\top}}  \cr
 		&=&  - \frac{1}{n} \sum_{i=1}^n  e^{-\hat{H}_i (\theta^*) } RV_i  \frac{\partial   \hat{H}_i (\theta^*) }{\partial \theta }  \frac{\partial    \hat{H}_i (\theta^*) }{ \partial \theta ^{\top}}  + o_p(1) \cr
 		&=&- \frac{1}{n} \sum_{i=1}^n  M_i  \frac{\partial   H_i (\theta_0) }{\partial \theta }  \frac{\partial    H_i (\theta_0) }{ \partial \theta ^{\top}} + o_p(1). 
 \end{eqnarray*}
 Since $IV_i$'s and $M_i$'s are non-degenerating, $\frac{1}{n} \sum_{i=1}^n  M_i  \frac{\partial   H_i (\theta_0) }{\partial \theta }  \frac{\partial    H_i (\theta_0) }{ \partial \theta ^{\top}}$ is positive definite. 
 Thus, by \eqref{eq1-thm-asym}, we have
 \begin{equation*}
 	\theta   - \theta_0= O_p ( n^{-1/2} + m^{-1/4}). 
 \end{equation*}

 Now, we show the asymptotic normality. 
 By Theorem \ref{thm-integratedVol}(b), we have 
\begin{eqnarray*}
	H_n (\theta) &=& \omega^g + (\gamma  + \beta ^g ) H_{n-1} (\theta) + \beta^g  \log M_{n-1} \cr
		&=&  \frac{ \omega^g}{ 1- \gamma - \beta^g }  +  \sum_{i=1}^{\infty}  \beta^g (\gamma  + \beta ^g ) ^ {i-1}  \log M_{n-i}.
\end{eqnarray*}
Since $\log M_i$'s are i.i.d., $(H_n (\theta), M_n) $ is strictly stationary.
By   Theorem 2.1 \citep{francq2013garch} and Theorem 2.5 \citep{bougerol1992strict}, $(H_n (\theta), M_n) $ is ergodic. 
 Then, applying the martingale central limit theorem, we obtain 
\begin{equation*}
\frac{1}{\sqrt{n}} \sum_{i=1}^n ( 1 - M_i ) \frac{\partial H_i (\theta_0) }{\partial \theta}   \overset{d} {\to} N(0, A  V).
\end{equation*}
By the ergodic theorem, we can show
\begin{equation*}
		-\triangledown \hat{s}_{n,m}(\theta^*) \overset{p}{\to} V .
\end{equation*}
Thus, by the Slutsky theorem, we have
\begin{equation*}
	\sqrt{n} ( \hat{\theta} - \theta_0) \overset{d} {\to} N(0,  A V^{-1}). 
\end{equation*}
\endpf

\section*{Acknowledgments} 
 
The authors appreciate the Editor, Professor D. Kristensen, and anonymous two referees for their careful reading of this paper and valuable comments.
The research of Donggyu Kim was supported in part by the National Research Foundation of Korea (NRF) (2021R1C1C1003216).


\bibliography{myReferences}

\begin{thebibliography}{}

\bibitem[A{\"\i}t-Sahalia et~al., 2010]{ait2010high}
A{\"\i}t-Sahalia, Y., Fan, J., and Xiu, D. (2010).
\newblock High-frequency covariance estimates with noisy and asynchronous
  financial data.
\newblock {\em Journal of the American Statistical Association},
  105(492):1504--1517.

\bibitem[A{\"\i}t-Sahalia and Xiu, 2016]{ait2016increased}
A{\"\i}t-Sahalia, Y. and Xiu, D. (2016).
\newblock Increased correlation among asset classes: Are volatility or jumps to
  blame, or both?
\newblock {\em Journal of Econometrics}, 194(2):205--219.

\bibitem[Andersen and Bollerslev, 1997a]{andersen1997heterogeneous}
Andersen, T.~G. and Bollerslev, T. (1997a).
\newblock Heterogeneous information arrivals and return volatility dynamics:
  Uncovering the long-run in high frequency returns.
\newblock {\em The journal of Finance}, 52(3):975--1005.

\bibitem[Andersen and Bollerslev, 1997b]{andersen1997intra-day}
Andersen, T.~G. and Bollerslev, T. (1997b).
\newblock Intraday periodicity and volatility persistence in financial markets.
\newblock {\em Journal of empirical finance}, 4(2-3):115--158.

\bibitem[Andersen and Bollerslev, 1998a]{andersen1998skeptics}
Andersen, T.~G. and Bollerslev, T. (1998a).
\newblock Answering the skeptics: Yes, standard volatility models do provide
  accurate forecasts.
\newblock {\em International Economic Review}, 39(4):885--905.

\bibitem[Andersen and Bollerslev, 1998b]{andersen1998deutsche}
Andersen, T.~G. and Bollerslev, T. (1998b).
\newblock Deutsche mark-dollar volatility: Intraday activity patterns,
  macroeconomic announcements, and longer run dependencies.
\newblock {\em The journal of Finance}, 53(1):219--265.

\bibitem[Andersen et~al., 2003]{andersen2003modeling}
Andersen, T.~G., Bollerslev, T., Diebold, F.~X., and Labys, P. (2003).
\newblock Modeling and forecasting realized volatility.
\newblock {\em Econometrica}, 71(2):579--625.

\bibitem[Andrews, 1992]{andrews1992generic}
Andrews, D.~W. (1992).
\newblock Generic uniform convergence.
\newblock {\em Econometric theory}, 8(2):241--257.

\bibitem[Barndorff-Nielsen et~al., 2008]{barndorff2008designing}
Barndorff-Nielsen, O.~E., Hansen, P.~R., Lunde, A., and Shephard, N. (2008).
\newblock Designing realized kernels to measure the ex post variation of equity
  prices in the presence of noise.
\newblock {\em Econometrica}, 76(6):1481--1536.

\bibitem[Bollerslev, 1986]{bollerslev1986generalized}
Bollerslev, T. (1986).
\newblock Generalized autoregressive conditional heteroskedasticity.
\newblock {\em Journal of econometrics}, 31(3):307--327.

\bibitem[Bougerol and Picard, 1992]{bougerol1992strict}
Bougerol, P. and Picard, N. (1992).
\newblock Strict stationarity of generalized autoregressive processes.
\newblock {\em The Annals of Probability}, 20(4):1714--1730.

\bibitem[Campbell and Thompson, 2008]{campbell2008predicting}
Campbell, J.~Y. and Thompson, S.~B. (2008).
\newblock Predicting excess stock returns out of sample: Can anything beat the
  historical average?
\newblock {\em The Review of Financial Studies}, 21(4):1509--1531.

\bibitem[Corsi, 2009]{corsi2009simple}
Corsi, F. (2009).
\newblock A simple approximate long-memory model of realized volatility.
\newblock {\em Journal of Financial Econometrics}, 7(2):174--196.

\bibitem[Diebold and Mariano, 2002]{diebold2002comparing}
Diebold, F.~X. and Mariano, R.~S. (2002).
\newblock Comparing predictive accuracy.
\newblock {\em Journal of Business \& economic statistics}, 20(1):134--144.

\bibitem[Engle, 1982]{engle1982autoregressive}
Engle, R.~F. (1982).
\newblock Autoregressive conditional heteroscedasticity with estimates of the
  variance of united kingdom inflation.
\newblock {\em Econometrica: Journal of the Econometric Society}, pages
  987--1007.

\bibitem[Fan and Kim, 2018]{fan2018robust}
Fan, J. and Kim, D. (2018).
\newblock Robust high-dimensional volatility matrix estimation for
  high-frequency factor model.
\newblock {\em Journal of the American Statistical Association},
  113(523):1268--1283.

\bibitem[Fan and Wang, 2007]{fan2007multi}
Fan, J. and Wang, Y. (2007).
\newblock Multi-scale jump and volatility analysis for high-frequency financial
  data.
\newblock {\em Journal of the American Statistical Association},
  102(480):1349--1362.

\bibitem[Francq et~al., 2013]{francq2013garch}
Francq, C., Wintenberger, O., and Zakoian, J.-M. (2013).
\newblock Garch models without positivity constraints: Exponential or log
  garch?
\newblock {\em Journal of Econometrics}, 177(1):34--46.

\bibitem[Hansen and Huang, 2016]{hansen2016exponential}
Hansen, P.~R. and Huang, Z. (2016).
\newblock Exponential garch modeling with realized measures of volatility.
\newblock {\em Journal of Business \& Economic Statistics}, 34(2):269--287.

\bibitem[Hansen et~al., 2012]{hansen2012realized}
Hansen, P.~R., Huang, Z., and Shek, H.~H. (2012).
\newblock Realized garch: a joint model for returns and realized measures of
  volatility.
\newblock {\em Journal of Applied Econometrics}, 27(6):877--906.

\bibitem[Jacod et~al., 2009]{jacod2009microstructure}
Jacod, J., Li, Y., Mykland, P.~A., Podolskij, M., and Vetter, M. (2009).
\newblock Microstructure noise in the continuous case: the pre-averaging
  approach.
\newblock {\em Stochastic processes and their applications}, 119(7):2249--2276.

\bibitem[Kawakatsu, 2006]{kawakatsu2006matrix}
Kawakatsu, H. (2006).
\newblock Matrix exponential garch.
\newblock {\em Journal of Econometrics}, 134(1):95--128.

\bibitem[Kim and Fan, 2019]{kim2019factor}
Kim, D. and Fan, J. (2019).
\newblock Factor garch-it{\^o} models for high-frequency data with application
  to large volatility matrix prediction.
\newblock {\em Journal of Econometrics}, 208(2):395--417.

\bibitem[Kim and Wang, 2016]{kim2016unified}
Kim, D. and Wang, Y. (2016).
\newblock Unified discrete-time and continuous-time models and statistical
  inferences for merged low-frequency and high-frequency financial data.
\newblock {\em Journal of Econometrics}, 194:220--230.

\bibitem[Kim et~al., 2016]{kim2016asymptotic}
Kim, D., Wang, Y., and Zou, J. (2016).
\newblock Asymptotic theory for large volatility matrix estimation based on
  high-frequency financial data.
\newblock {\em Stochastic Processes and their Applications}, 126:3527–--3577.

\bibitem[Nelson, 1991]{nelson1991conditional}
Nelson, D.~B. (1991).
\newblock Conditional heteroskedasticity in asset returns: A new approach.
\newblock {\em Econometrica: Journal of the Econometric Society}, pages
  347--370.

\bibitem[Shephard and Sheppard, 2010]{shephard2010realising}
Shephard, N. and Sheppard, K. (2010).
\newblock Realising the future: forecasting with high-frequency-based
  volatility (heavy) models.
\newblock {\em Journal of Applied Econometrics}, 25(2):197--231.

\bibitem[Shin et~al., 2021]{shin2021adaptive}
Shin, M., Kim, D., and Fan, J. (2021).
\newblock Adaptive robust large volatility matrix estimation based on
  high-frequency financial data.
\newblock {\em Available at SSRN 3793394}.

\bibitem[Song et~al., 2021]{song2021volatility}
Song, X., Kim, D., Yuan, H., Cui, X., Lu, Z., Zhou, Y., and Wang, Y. (2021).
\newblock Volatility analysis with realized garch-it{\^o} models.
\newblock {\em Journal of Econometrics}, 222(1):393--410.

\bibitem[Tao et~al., 2011]{tao2011large}
Tao, M., Wang, Y., Yao, Q., and Zou, J. (2011).
\newblock Large volatility matrix inference via combining low-frequency and
  high-frequency approaches.
\newblock {\em Journal of the American Statistical Association},
  106(495):1025--1040.

\bibitem[Xiu, 2010]{xiu2010quasi}
Xiu, D. (2010).
\newblock Quasi-maximum likelihood estimation of volatility with high frequency
  data.
\newblock {\em Journal of Econometrics}, 159(1):235--250.

\bibitem[Zhang, 2006]{zhang2006efficient}
Zhang, L. (2006).
\newblock Efficient estimation of stochastic volatility using noisy
  observations: A multi-scale approach.
\newblock {\em Bernoulli}, 12(6):1019--1043.

\bibitem[Zhang et~al., 2005]{zhang2005tale}
Zhang, L., Mykland, P.~A., and A{\"\i}t-Sahalia, Y. (2005).
\newblock A tale of two time scales: Determining integrated volatility with
  noisy high-frequency data.
\newblock {\em Journal of the American Statistical Association},
  100(472):1394--1411.

\bibitem[Zhang et~al., 2016]{zhang2016jump}
Zhang, X., Kim, D., and Wang, Y. (2016).
\newblock Jump variation estimation with noisy high frequency financial data
  via wavelets.
\newblock {\em Econometrics}, 4(3):34.

\end{thebibliography}
\end{spacing}
\end{document}